\documentclass[a4paper,11pt]{article}
\usepackage{amsmath,amssymb,amsthm,amsfonts}    % AMS math packages
\usepackage{mathrsfs,bm}                        % Bold math symbols
\usepackage{dsfont}
\usepackage{eurosym}                            % Euro symbol
\usepackage{xcolor}                             % Color text management
\usepackage{url,hyperref}                       % Urls and hypertexts (for online files) management
\usepackage{float}                              % Floating objects management
\usepackage[small,bf]{caption}                  % Advanced caption management
\usepackage{graphicx}
\usepackage{epstopdf}                           % Figures management
\usepackage{subfigure}
\usepackage{psfrag}                             % Allows to overlay EPS figures with arbitrary LaTeX constructions
\usepackage{epsfig}                             % Encapsulated postscript management
\graphicspath{{./Figures/}}                     % Figures path.
\usepackage{booktabs,multirow,longtable}        % Advanced table management
\usepackage{tabularx}        % Advanced table management
\usepackage[toc,page]{appendix}
\usepackage[english]{babel}
\usepackage{mathtools}
\usepackage{makecell}
\usepackage[margin=0.5in]{geometry}
\usepackage{booktabs}
\usepackage[nottoc,notlot,notlof]{tocbibind}
\makeatletter \@addtoreset{equation}{section} \makeatother

\begin{document}
% --- Definitions of new Latex commands ---
%
% REMIND: always keep a single file with all definitions, beware of multiple copies.

% --- Text ---
\newcommand{\ie}{i.e.~} 	                    % i.e. with non-breakable space
\newcommand{\eg}{e.g.~} 	                    % e.g. with non-breakable space
\newcommand{\rhs}{r.h.s.~} 	                    % r.h.s. with non-breakable space
\newcommand{\wrt}{w.r.t.~} 	                    % w.r.t. with non-breakable space
\newcommand{\old}[1]{\textcolor{blue}{#1}}      % old text, blue, strike-through tex
\newcommand{\new}[1]{\textcolor{red}{#1}}       % new text, red, strike-through tex
\newcommand{\comment}[1]{\textcolor{green}{#1}} % comments, green, strike-through tex
\newcommand{\textbox}[1]{\mbox{\textit{#1}}} 	% text inside math

% --- Mathematical Definitions ---
\newtheorem{theorem}{Theorem}[section]
\newtheorem{definition}[theorem]{Definition}
\newtheorem{definitions}[theorem]{Definition}
\newtheorem{proposition}[theorem]{Proposition}
\newtheorem{remark}[theorem]{Remark}
\newtheorem{corollary}[theorem]{Corollary}
\newtheorem{example}[theorem]{Example}
\newtheorem{lemma}[theorem]{Lemma}

% --- Environments ---
\newenvironment{system}{\left\lbrace\begin{array}{@{}l@{}}}{\end{array}\right.}

% --- Delimiters ---
\newcommand{\Parenthesis}[1]{\left( #1 \right)}         % (xxx)
\newcommand{\Brack}[1]{\left\lbrack #1 \right\rbrack}   % [xxx]
\newcommand{\Brace}[1]{\left\lbrace #1 \right\rbrace}   % {xxx}
\newcommand{\Abs}[1]{\left\lvert #1 \right\rvert}       % |xxx|
\newcommand{\Norm}[1]{\left\lVert #1 \right\rVert}      % ||xxx||
\newcommand{\Mean}[1]{\left\langle #1 \right\rangle}    % <xxx>
\newcommand{\QuoteDouble}[1]{``#1''}                    % double quotation marks
\newcommand{\QuoteSingle}[1]{`#1'}                      % single quotation marks

% --- Math operators ---
\newcommand{\beq}{\begin{equation}}         % \begin{equation}
\newcommand{\eeq}{\end{equation}}           % \end{equation}
\newcommand{\Max}{\mbox{Max}}                                   % Max operator
\newcommand{\Min}{\mbox{Min}}                                   % Min operator
\newcommand{\Derp}[2]{\frac{\partial #1}{\partial #2}}          % partial derivative, 1st order
\newcommand{\DerpXX}[2]{\frac{\partial^2 #1}{\partial #2 ^2}}   % partial derivative, 2nd order
\newcommand{\DerpXY}[3]{\frac{\partial^2 #1}{\partial #2 \partial #3}}    % partial derivative, 2nd order, cross
\newcommand{\half}[0]{\frac{1}{2}}                                                  % one half
\newcommand{\ind}[1]{\mathbf{1}_{#1}} 	                                            % indicator
\newcommand{\E}[1]{\mathbb{E}\left[#1\right]}                                       % Expected value
\newcommand{\expt}[1]{\mathbb{E}_t\left[#1\right]}                                  % Expected value at time t
\newcommand{\Econd}[2]{\mathbb{E}\left[#1\;\middle \vert\;\mathcal{F}_{#2}\right]}  % Expected value conditioned

% --- Sets ---
\newcommand{\Nset}{\mathbb{N}}      % Natural numbers set
\newcommand{\Zset}{\mathbb{Z}}      % Integer numbers set
\newcommand{\Qset}{\mathbb{Q}}      % Rational numbers set
\newcommand{\Rset}{\mathbb{R}}      % Real numbers set

% --- Finance ---
\newcommand{\Black}{\mbox{Black}}				    % Black's formula
\newcommand{\XVA}{\mbox{\textit{XVA}}}				% XVA
\newcommand{\CVA}{\mbox{\textit{CVA}}}				% CVA
\newcommand{\DVA}{\mbox{\textit{DVA}}}				% DVA
\newcommand{\FVA}{\mbox{\textit{FVA}}}				% FVA

% --- Financial Instruments ---
\newcommand{\Depo}{\mbox{\textbf{Depo}}}			% Deposit
\newcommand{\FRA}{\mbox{\textbf{FRA}}}			    % Forward Rate Agreement
\newcommand{\Futures}{\mbox{\textbf{Futures}}}      % Futures
\newcommand{\ZCB}{\mbox{ZCB}}					    % Zero Coupon Bond
\newcommand{\Swap}{\mbox{\textbf{Swap}}}			% Swap
\newcommand{\Swaplet}{\mbox{\textbf{Swaplet}}}		% Swaplet
\newcommand{\IRS}{\mbox{\textbf{IRS}}}			    % Interest Rate Swap
\newcommand{\IRSlet}{\mbox{\textbf{IRSlet}}}        % Interest Rate Swaplet
\newcommand{\OIS}{\mbox{\textbf{OIS}}}              % OIS (Overnight Indexed Swap)
\newcommand{\OISlet}{\mbox{\textbf{OISlet}}}        % OISlet
\newcommand{\BSwap}{\mbox{\textbf{BSwap}}}          % Basis Swap
\newcommand{\BSwaplet}{\mbox{\textbf{BSwaplet}}}    % Basis Swaplet
\newcommand{\IRBS}{\mbox{\textbf{IRBS}}}            % Interest Rate Basis Swap
\newcommand{\IRBSlet}{\mbox{\textbf{IRBSlet}}}      % Interest Rate Basis Swaplet
\newcommand{\CCSwap}{\mbox{\textbf{CCSwap}}}        % Cross Currency Swap
\newcommand{\CCS}{\mbox{\textbf{CCS}}}              % Cross Currency Swap
\newcommand{\CCSwaplet}{\mbox{\textbf{CCSwaplet}}}  % Cross Currency Swaplet
\newcommand{\CCSlet}{\mbox{\textbf{CCSlet}}}        % Cross Currency Swaplet
\newcommand{\Caplet}{\mbox{\textbf{Caplet}}}        % Caplet
\newcommand{\Floorlet}{\mbox{\textbf{Floorlet}}}    % Floorlet
\newcommand{\cf}{\mbox{\textbf{cf}}}                % Caplet/Floorlet
\newcommand{\CAP}{\mbox{\textbf{Cap}}}              % Cap
\newcommand{\Floor}{\mbox{\textbf{Floor}}}          % Floor
\newcommand{\CF}{\mbox{\textbf{CF}}}                % Cap/Floor
\newcommand{\Swaption}{\mbox{\textbf{Swaption}}}    % Swaption
\newcommand{\CMSlet}{\mbox{\textbf{CMSlet}}}        % CMSlet
\newcommand{\CMS}{\mbox{\textbf{CMS}}}              % CMS
\newcommand{\CMScf}{\mbox{\textbf{CMScf}}}          % CMS caplet/floorlet
\newcommand{\FXFwd}{\mbox{\textbf{FXFwd}}}          % Forex forward
\title{\textbf{No Fear of Discounting\\How to Manage the Transition from EONIA to \euro STR}}

\author{
	Marco Scaringi\footnote{Intesa Sanpaolo, Financial and Market Risk Management, \texttt{marco.scaringi@intesasanpaolo.com}},
	Marco Bianchetti\footnote{Intesa Sanpaolo, Financial and Market Risk Management, and University of Bologna, Department of Statistical Sciences \QuoteDouble{Paolo Fortunati}, \texttt{marco.bianchetti@unibo.it}}
}

\date{29 January 2021}
\maketitle

\begin{abstract}
	An important step in the Financial Benchmarks Reform \cite{FSB14} was taken on 13th September 2018, when the ECB Working Group on Euro Risk-Free Rates recommended the Euro Short-Term Rate \euro STR as the new benchmark rate for the euro area, to replace the Euro OverNight Index Average (EONIA) which will be discontinued at the end of 2021. This transition has a number of important consequences on financial instruments, OTC derivatives in particular. 
	\par 
	In this paper we show in detail how the switch from EONIA to \euro STR affects the pricing of OIS, IRS and XVAs. We conclude that the adoption of the \QuoteDouble{clean discounting} approach recommended by the ECB \cite{ECB19a}, based on \euro STR only, is theoretically sound and leads to very limited impacts on financial valuations. 
	\par 
	This finding ensures the possibility, for the financial industry, to switch all EUR OTC derivatives, either cleared with Central Counterparties, or subject to bilateral collateral agreements, or non-collateralised, in a safe and consistent manner. The transition to such EONIA-free pricing framework is essential for the complete elimination of EONIA before its discontinuation scheduled on 31st December 2021. 
\end{abstract}

\newpage
\tableofcontents

\vspace{2cm} 

\vspace{0.5cm} 

\vspace{0.5cm} 
\noindent \textbf{Acknowledgements}: The authors acknowledge fruitful discussions with L. Cefis, F. Fogliani, N. Moreni, and many other colleagues in Intesa Sanpaolo.

\vspace{0.5cm} 
\noindent \textbf{Disclaimer}: the views expressed here are those of the authors and do not represent the opinions of their employers. They are not responsible for any use that may be made of these contents.

\newpage

%%%%%%%%%%%%%%%%%%%%%%%%%%%%%%%%%%%%%%%%%%%

\section{Introduction}
\label{Sec: Intro}
The Financial Benchmarks Reform aims to strengthen the reliability of the most important interest rates in front of the weaknesses observed after the credit crunch crisis of 2007. 
In February 2013 the G20 gave mandate to the Financial Stability Board (FSB) to promote a reform process of the principal financial benchmarks. In July 2014 the FSB issued two important recommendations: a) to strengthen Interbank Offered rates (IBORs), in particular linking the fixing procedure to real transactions, and b) the identification of risk-free rates (RFR) alternative to IBORs \cite{FSB14}.
\par 
Focusing on the European Union (EU), the EU Benchmark Regulation, published on 8th June 2016 and expected to enter into force on 1st January 2022 \cite{BMR16}, sets the new rules regarding financial benchmarks.
The European Money Market Institute (EMMI), the administrator of the Euro OverNight Index Average (EONIA) \cite{EONIA} and of the Euro Interbank Offered rate (EURIBOR) \cite{EURIBOR}, launched in 2016 a review programme in order to align these benchmark rates with the requirements of the BMR. 
While EURIBOR was successfully reformed and declared BMR-compliant on 3rd July 2019, both EMMI's and ECB's extensive data analyses showed that EONIA could not be reformed, and EMMI communicated to stop any further effort on 1st February 2018. As a consequence, EONIA is expected to be discontinued on 31st December 2021 (the last EONIA fixing will be published on 3rd January 2022).
\par
On 26th January 2018 the European Central Bank (ECB) established the Working Group on Euro Risk-Free Rates \cite{WGERFR} in order to identify and recommend new risk-free rates alternative to the current benchmarks used in contracts. 
On 13th September 2018 the Working Group recommended the euro short-term rate \euro STR as the new risk-free rate for the euro area. 
The \euro STR reflects the wholesale euro unsecured overnight borrowing costs of euro area banks, it is published on each TARGET2 business day \QuoteDouble{T+1} based on transactions conducted and settled on the previous TARGET2 business day (reporting date “T”) with a maturity date of T+1 \cite{ECBESTR,ECBester}.
Since 2nd October 2019 EONIA is published by the ECB as \euro STR plus a fixed spread equal to $8.5$ basis points. 
Such spread was calculated by the ECB on 31st May 2019 as the average difference between EONIA and the pre-\euro STR in the period from 17th April 2018 to 16th April 2019.
\par
Focusing on derivative instruments, the introduction of \euro STR and the EONIA discontinuation has a number of important consequences:
\begin{enumerate}
	\item the OTC derivatives' market has to switch to trade \euro STR-indexed instruments, i.e. \euro STR Futures, \euro STR Forward Rate Agreements (FRAs), \euro STR Overnight Indexed Swaps (OISs), instead of the corresponding EONIA-linked instruments, and possibly to start new \euro STR-linked options, i.e. \euro STR Caps/Floors, \euro STR Swaptions;
	\item the legacy EONIA-linked instruments with maturity beyond the EONIA discontinuation date, i.e. EONIA OISs\footnote{other EONIA-linked derivatives, i.e. EONIA Futures and FRAs, typically have short maturities which do not cross the EONIA discontinuation date.}, has to be converted to \euro STR using the fall-back protocol elaborated by ISDA\footnote{On 1st October 2019 ISDA published supplements 59 and 60 to the 2006 ISDA Definitions, which provide a new Floating Rate Option for \euro STR and an amended version of the EONIA Floating Rate Options, so that they have fallbacks based on the EU Risk Free Rate Working Group’s recommendation. See also \url{https://www.isda.org/2020/05/11/benchmark-reform-and-transition-from-libor} (URL visited on 24th July 2020).
	};
	\item the discounting rate and the Price Alignment (PAI) rate used by Central Counterparties (CCPs) to discount future cash flows of cleared EUR trades and to compute EUR collateral interest amounts has to be switched from EONIA to \euro STR\footnote{LCH and EUREX planned the discounting switch on 27th July 2020, using cash compensation or dummy trades to manage the corresponding NPV and collateral jumps. See  \url{https://www.lch.com/membership/ltd-membership/ltd-member-updates/transition-to-\%E2\%82\%ACSTR-Discounting-Updated-Timing} for LCH and \url{https://www.eurexgroup.com/group-en/newsroom/circulars/clearing-circular-1942440} for EUREX. URLs visited on 24th July 2020.}; 
	\item consistently with the CCPs' discounting switch, also bilateral collateral agreements covering non-cleared EUR OTC derivatives using EONIA collateral rate should be re-negotiated to \euro STR;
	\item finally, consistently with cleared and collateralized derivatives, also non-collateralized EUR OTC derivatives where EONIA is used as discounting rate should be switched to \euro STR-discounting;
	\item as a consequence of the changes above, and in particular of the CCPs EUR discounting switch, market quotes of traded EUR OTC derivatives has to be interpreted as cleared or collateralized using \euro STR for discounting and margination, and the associated implicit quantities, e.g. forwards, volatilities, correlations, etc. have to be derived using \euro STR-discounting. In particular, EUR single- and multi-currency yield curves have to be bootstrapped from EUR Overnight Indexed Swaps (OISs), Interest Rate Swaps (IRSs) and Cross Currency Swaps (CCSs) using \euro STR-discounting.
\end{enumerate}
\par 
While the CCPs' switch was decided centrally after extensive consultations, the bilateral CSA switch is left to direct agreements between conterparties. No-arbitrage requires that trades under bilateral CSA are priced consistently with the collateral rate specified in the CSA. If the collateral rate is different from \euro STR, the price differs from the corresponding \euro STR-discounting price, the difference being called Collateral Valuation Adjustment (COLVA, see \cite{BriPal13,Green15,Greg20}). Clearly, managing multiple discounting regimes for different netting sets, CSAs and counterparties creates an additional operational complexity. Because of this reasons the Working Group on Euro Risk-Free Rates encouraged \QuoteDouble{\emph{market participants to make all reasonable efforts to replace EONIA with the \euro STR as a basis for collateral interest for both legacy and new trades with each of its counterparties (clean discounting)}} \cite{ECB19a}.
\par 
Still to avoid arbitrage possibilities, it is natural to adopt \euro STR-discounting also for any other non-collateralized OTC derivative, but in this case also the associated valuation adjustments (XVAs, see e.g. \cite{BriPal13,Green15,Greg20}) have to be switched consistently, including, in particular, the funding spread used for Funding Valuation Adjustment (FVA). 
\par
In this paper we show in detail how the switch from EONIA to \euro STR overnight rates affects the pricing of OIS, IRS and their associated XVAs, and what are the consequences on cleared, collateralised and non-collateralised linear financial instruments. 
The paper is organized as follows: in sec. \ref{Sec: Basics} we introduce the notation, we briefly remind the basic theoretical framework, we proof the equivalence of EONIA and \euro STR forward probability measures, and we remind the OIS, IRS and FVA formulas; in sec. \ref{Sec: OIS impacts} we analyse the theoretical and numerical impacts of the transition on OISs, both in the true discrete and in the approximated continuous compounding regimes; in sec. \ref{Sec: IRS impacts} we analyse IRS; in sec. \ref{Sec: XVA impacts} we analyse the impacts of the transition including XVAs, while in sec. \ref{Sec: conclusion} we draw the conclusions.

\section{Preliminaries}
\label{Sec: Basics}

\subsection{Theoretical Framework}
\label{Subsec: theory}
One of the major innovations in financial mathematics after the credit crunch crisis was that, assuming no arbitrage and the usual probabilistic framework ($\Omega,\mathcal{F},\mathcal{F}_t,\mathbb{Q}$) with market filtration $\mathcal{F}_{t}$ and risk-neutral probability measure $\mathbb{Q}$, the general pricing formula of a financial instrument with payoff $V(T)$ paid at time $T>t$ is
\begin{align}
V^\alpha(t) &= V^\alpha_{0}(t) + \text{XVA}^\alpha(t)\label{eq: fair value},\\
V^\alpha_{0}(t) &=\mathbb{E}^{\mathbb{Q}}\Brack{D^\alpha(t;T)V(T)\left|\mathcal{F}_t\right.}
=P^\alpha(t;T) \mathbb{E}^{\mathbb{Q}^\alpha_T}\Brack{V(T)\left|\mathcal{F}_t\right.},\label{eq: base value}\\
D^\alpha(t;T) &= \dfrac{B^\alpha(t)}{B^\alpha(T)}=e^{-\int_{t}^{T}r_\alpha(u)du},\label{eq: stochastic discount factor}\\
P^\alpha(t;T) &=\mathbb{E}^{\mathbb{Q}}\Brack{D^\alpha(t;T)\left|\mathcal{F}_t\right.},\label{eq: ZCBrf}
\end{align}
where the base value (sometimes also called mark to market) $V^\alpha_{0}(t)$ in eqs. \eqref{eq: fair value}-\eqref{eq: base value} is interpreted as the price of the financial instrument under perfect $\alpha-$collateralization\footnote{i.e. an ideal CSA ensuring a perfect match between the price $V_{0}(t)$ and the corresponding variation margin $C(t)$ at any time $t$. This condition is realised in practice with a real CSA minimizing any friction between the price and the collateral, i.e. with daily margination, cash collateral in the same currency of the trade, flat overnight collateral rate, zero threshold, minimum transfer amount and independent amount.}, 
the discount (short) rate $r_\alpha(t)$ in eq. \eqref{eq: stochastic discount factor} is the corresponding $\alpha-$collateral rate, $B^\alpha(t)$ is the collateral bank account growing at rate $r_\alpha(t)$, $D^\alpha(t;T)$ is the stochastic collateral discount factor, $P^\alpha(t;T)$ is the collateralised Zero Coupon Bond (ZCB) price, and $\mathbb{Q}^\alpha_T$ is the $T$-forward probability measure associated to the numeraire $P^\alpha(t;T)$. 
We stress that the same instrument may be subject to different CSAs with different collateral rates $r_\alpha,r_\beta$, etc. Hence different discount factors, Zero Coupon Bonds and $T-$forward measures may be used in eqs. \eqref{eq: base value}-\eqref{eq: ZCBrf}, leading to different prices $V^\alpha_0(t)$. 
This is the reason why we specified the $\alpha-$collateral everywhere. 
We also stress that the $T$-forward measures $\mathbb{Q}^\alpha_T$ are different, but the risk neutral measure $\mathbb{Q}$ is unique, as pointed out in the literature (see e.g. \cite{Pit12,Bri18}). 
We will denote with subscripts the rate characteristics, e.g. the rate typology, and with apices the collateral characteristics.
The notation above is helpful since we have to deal with EONIA and \euro STR both as underlying and collateral rates. 
Tab. \ref{table: domains} summarizes the pricing of the base component of fair value in both discounting frameworks. 
\begin{table}[H]
	\centering
	\begin{tabularx}{\textwidth}{Xcc}
		\toprule
		Quantity						& EONIA				& \euro STR	\\
		\midrule
		Collateral rate 				& $r_\textit{EON}$  & $r_\textit{\euro ST}$ \\
		Risk-neutral measure pricing 	& $V_0^\textit{EON}(t) = \mathbb{E}_t^{\mathbb{Q}}\Brack{D^\textit{EON}(t;T)V(T)}$  & $V_0^\textit{\euro ST}(t) = \mathbb{E}_t^{\mathbb{Q}}\Brack{D^\textit{\euro ST}(t;T)V(T)}$\\
		T-Forward measure pricing 		& $V_0^\textit{EON}(t) = P^\textit{EON}(t;T)\mathbb{E}_t^{\mathbb{Q}_{T}^\textit{EON}}\Brack{V(T)}$ & $V_0^\textit{\euro ST}(t) = P^\textit{\euro ST}(t;T)\mathbb{E}_t^{\mathbb{Q}_{T}^\textit{\euro ST}}\Brack{V(T)}$ \\
		Stochastic discount factor 		& $D^\textit{EON}(t;T)=e^{-\int_{t}^{T}r_\textit{EON}(u)du}$ & $D^\textit{\euro ST}(t;T)=e^{-\int_{t}^{T}r_\textit{\euro ST}(u)du}$ \\
		Zero Coupon Bond 				& $P^\textit{EON}(t;T)=\mathbb{E}_t^{\mathbb{Q}}\Brack{D^\textit{EON}(t;T)}$  & $P^\textit{\euro ST}(t;T)=\mathbb{E}_t^{\mathbb{Q}}\Brack{D^\textit{\euro ST}(t;T)}$\\
		EONIA-\euro STR spread			&\multicolumn{2}{c}{$\Delta = r_{\textit{\euro ST}}(t) - r_{\textit{EON}}(t) = -8.5$ bps.} \\
		\bottomrule	     
	\end{tabularx}
	\caption{EONIA and \euro STR pricing frameworks for a derivative with payoff $V(T)$ at maturity $T$. We denote with $\alpha\in\Brace{\textit{EON,\euro ST}}$ the EONIA and \euro STR related quantities, respectively. To simplify the notation, we omit filtrations explicited e.g. in eq. \eqref{eq: base value}.}
	\label{table: domains}
\end{table}
\par 
Valuation adjustments in eq. \eqref{eq: fair value}, collectively named XVA, represent a crucial and consolidated component in modern derivatives pricing which takes into account additional risk factors not included among the risk factors considered in the base value $V_0$ in eq. \eqref{eq: base value}. 
These risk factors are typically related to counterparties default, funding, and capital, leading, respectively, to Credit/Debt Valuation Adjustment (CVA/DVA), Funding Valuation Adjustment (FVA), often split into Funding Cost/Benefit Adjustment (FCA/FBA) and initial Margin Valuation Adjustment (MVA), Capital Valuation Adjustment (KVA). A complete discussion on XVAs may be found e.g. in \cite{Green15,Greg20}. 
Hence, for XVAs pricing we must consider the enlarged filtration 
$ \mathcal{G}_{t}:= \mathcal{F}_{t}\vee\mathcal{H}_{t}\supseteq\mathcal{F}_{t}$ 
where $ \mathcal{H}_{t}=\sigma(\{\tau\leq u\}:u\leq t)$ 
is the filtration generated by default events (see e.g. \cite{BriMer06}). 
More details on XVAs pricing are discussed in sec. \ref{Subsec: XVA} below. 
Notice that we do not consider here the Additional Valuation Adjustments (AVAs) mentioned in the Basel Framework\footnote{\url{https://www.bis.org/basel_framework}, URL visited on 13th Aug. 2020.} and in the EU Capital Requirement Regulation (CRR, \cite{EUParlPrd13}), since they are not accounted at fair value but through capital (CET1).
\par 
The theoretical setting above is supported by an extensive literature, see e.g. \cite{Pit10a,Pit12,Bri18} and refs. therein.

\subsection{Equivalence of Forward Pricing Measures}
\label{Subsec: measure equiv}
In this section we show that the EONIA and \euro STR forward probability measures $\mathbb{Q}_T^\textit{EON}$ and $\mathbb{Q}_T^\textit{\euro ST}$ are equivalent. 
First of all we write the relationship between $V_0^{\textit{EON}}(t)$ and $V^{\textit{\euro ST}}_0(t)$ as follows,
\begin{equation}
\begin{split}
V_{0}^{\textit{\euro ST}}(t)&=\mathbb{E}_t^{\mathbb{Q}}\Brack{D^{\textit{\euro ST}}(t;T)\,V(T)}\\
&=\mathbb{E}_t^{\mathbb{Q}}\Brack{e^{-\int_{t}^{T}\Brack{r_{\textit{EON}}(u)+\Delta}\,du}\,V(T)}\\
%&=e^{-\int_{t}^{T}\Delta du}\,\mathbb{E}_t^{\mathbb{Q}}\Brack{D(t;T)V_{0}(T)}\\
&=e^{-\Delta\,\tau(t,T)}V_{0}^{\textit{EON}}(t).
\label{eq: Vest}
\end{split}
\end{equation}
For ZCBs a similar relationship holds,
\begin{equation}\begin{split}
P^{\textit{\euro ST}}(t;T)&=\mathbb{E}_t^{\mathbb{Q}}\Brack{D^{\textit{\euro ST}}(t;T)}\\
&=\mathbb{E}_t^{\mathbb{Q}}\Brack{e^{-\int_{t}^{T}\Parenthesis{r_{\textit{EON}}(u)+\Delta}\,du}}\\
%&=e^{-\int_{t}^{T}\Delta du}\mathbb{E}_t^{\mathbb{Q}}\Brack{e^{-\int_{t}^{T}r(u)du}}\\
&=e^{-\Delta\tau(t,T)}P^{\textit{EON}}(t;T).
\label{eq: Pest}
\end{split}\end{equation}
Eqs. \eqref{eq: Vest} and \eqref{eq: Pest} lead to
\beq
\mathbb{E}_t^{\mathbb{Q}_{T}^{\textit{\euro ST}}}\Brack{V(T)}
=\dfrac{V_{0}^{\textit{\euro ST}}(t)}{P^{\textit{\euro ST}}(t;T)}
=\dfrac{e^{-\Delta\tau(t,T)}V_{0}^{\textit{EON}}(t)}{e^{-\Delta\tau(t,T)}P^{\textit{EON}}(t;T)} 
=\dfrac{V_{0}^{\textit{EON}}(t)}{P^{\textit{EON}}(t;T)}
=\mathbb{E}_t^{\mathbb{Q}_{T}^{\textit{EON}}}\Brack{V(T)}.
\label{eq: measure equiv}
\eeq
Equation \eqref{eq: measure equiv} proves the equivalence of the two forward measures $\mathbb{Q}_T^{\textit{EON}}$ and $\mathbb{Q}^{\textit{\euro ST}}_T$. 
A similar derivation can be found in \cite{ECB19c}. 
We notice that this property holds for any deterministic non-costant spread and even for a stochastic spread independent of $r_{\textit{EON}}$.

\subsection{OIS and IRS Pricing}
\label{Subsec: OIS and IRS pricing}
In this section we remember the pricing formulas for Interest Rate Swaps (IRSs) and Overnight Indexed Swaps (OISs), using a single unified notation enconpassing both cases.
\par 
We consider a generic Swap contract, which allows the exchange of a fixed rate $K$ against a floating rate, characterised by the following time schedules and pyoffs for the floating and fixed legs, respectively,
\begin{equation}
\begin{split}
&\bm{T}=[T_0,...,T_i,...,T_n],\quad\textit{floating leg}\\
&\bm{S}=[S_0,...,S_j,...,S_m],\quad\textit{fixed leg}\\
&T_0=S_0,\quad T_n = S_m,\\
&\Swaplet_\textit{float}(T_i;T_{i-1},T_i) = N R_x(T_{i-1},T_i) \tau_R(T_{i-1},T_i),\\
&\Swaplet_\textit{fix}(S_j;S_{j-1},S_j,K) = N K \tau_K(S_{j-1},S_j),
\end{split}
\end{equation}
where $\tau_K$ and $\tau_R$ are the year fractions for fixed and floating rate conventions, respectively, and $R_x(T_{i-1},T_i)$ is the underlying spot floating rate with tenor $x$, consistent with the time interval $\Brack{T_{i-1},T_i}$, as explained below.   
\par 
The Swap's price at time $t$ is given by the sum of the prices of fixed and floating cash flows occurring after $t$, 
\begin{align}
\Swap(t;\bm{T},\bm{S},K,\omega,\alpha)
&= \omega N \Brack{\sum_{i=\eta_R(t)}^n P^\alpha(t;T_i) F^\alpha_{x,i}(t) \tau_R(T_{i-1},T_i) - K A(t;\bm{S},\alpha)},\label{eq: IRS price}\\
A(t;\textbf{S},\alpha) &= \sum_{j=\eta_K(t)}^m P^\alpha(t;S_j)\tau_K(S_{j-1},S_j),
\label{eq: IRS annuity}
\end{align}
where $N$ is a nominal amount, $\omega =+/- 1 $ for a payer/receiver Swap, 
$\eta_R(t)=\min\{i\in\{1,...,n\}\text{ s.t. } T_i\geq t\}$ and 
$\eta_K(t)=\min\{j\in\{1,...,m\}\text{ s.t. } S_j\geq t\}$ indexes the first future cash flows in the Swap's schedules, $A(t;\textbf{S},\alpha)$ is the Swap's annuity, $\alpha$ denotes the Swap's CSA with collateral rate $r_\alpha$, and $F^\alpha_{x,i}(t)$ is the forward rate observed at time $t$, fixing at future time\footnote{if $T_{i-1}\leq t \leq T_i$ the rate has already fixed, hence $F^\alpha_{x,i}(t)=R_{x}(T_{i-1},T_i)$.} $T_{i-1}$ and spanning the future time interval $\Brack{T_{i-1},T_{i}}$, given by
\begin{equation}
F^\alpha_{x,i}(t):=F^\alpha_x(t;T_{i-1},T_i)
:=\mathbb{E}^{\mathbb{Q}^\alpha_{T_i}}_{t}\Brack{R_x(T_{i-1},T_i)}.
%=\frac{1}{\tau_R(T_{i-1},T_i)}\left[\frac{P_x\left(t;T_{i-1}\right)}{P_x\left(t;T_i\right)}-1\right].
\label{eq: IRS forward}
\end{equation}
By construction, the forward rate $F^\alpha_{x,i}(t)$ is a martingale under the forward measure $Q^\alpha_{T_{i}}$ associated to the CSA-numeraire $P^\alpha(t;T_i)$. 
\par 
The par swap rate $R_x(t;\textbf{T},\textbf{S},\alpha)$, i.e. the fixed rate $K$ such that the Swap is worth zero, is given by
\begin{equation}
R_x(t;\textbf{T},\textbf{S},\alpha)= \dfrac{\sum_{i=\eta_R(t)}^n P^\alpha(t;T_i)F^\alpha_{x,i}(t) \tau_R(T_{i-1},T_i)}{A(t;\textbf{S},\alpha)}.
\label{eq: IRS par rate}
\end{equation}
\par 
In the case of IRS the underlying rate $R_x(T_{i-1},T_i)$ is an IBOR with a tenor $x$ consistent with the time interval $\Brack{T_{i-1},T_i}$ (e.g. $x=6M$ for EURIBOR 6M and semi-annual coupons). 
IBOR forward rates $F^\alpha_{x,i}(t)$ are computed from IBOR yield curves built from homogeneous market IRS quotes (i.e. with the same underlying IBOR tenor $x$), using the corresponding OIS yield curve for discounting, a procedure commonly called multi-curve bootstrapping\footnote{since OIS and IBOR curves with different tenors are involved, see e.g. \cite{AmeBia13,Hen14} for a detailed discussion.}. In the present context we must also stress that both market IRS and OIS used for yield curve bootstrapping must share the same $\alpha-$CSA. 
Only in this case we may write the usual expression of forward rates
\begin{equation}
F^\alpha_{x,i}(t)
=\frac{1}{\tau_F(T_{i-1},T_i)}\left[\frac{P^\alpha_x(t;T_{i-1})}{P^\alpha_x(t;T_i)}-1\right],
\label{eq: IBOR forward rate vs risky ZCBs}
\end{equation}
where $\tau_F$ is the year fraction with the forward rate convention and $P^\alpha_x(t;T_i)$ can be interpreted as the price of a risky ZCB issued by an average IBOR counterparty\footnote{i.e. an issuer with a credit risk equal to the average credit risk of the IBOR panel, see e.g. \cite{Mor09} and App. \ref{App: riskyZCB}. We stress that the $\alpha-$CSA index in $P^\alpha_x(t;T_i)$ is simply inherited from the forward rate $F^\alpha_{x,i}(t)$ on the l.h.s. of eq. \eqref{eq: IBOR forward rate vs risky ZCBs}, which, according to eq. \eqref{eq: IRS forward}, is associated to an instrument with $\alpha-$CSA, but it does not refer to a CSA associated to risky ZCBs.}. We stress that eq. \eqref{eq: IBOR forward rate vs risky ZCBs} above is a recursive definition of $P^\alpha_x(t;T_i)$ at $T_i$, given the market forward rate $F^\alpha_{x,i}(t)$ at $\Brack{T_{i-1},T_i}$ and $P^\alpha_x(t;T_{i-1})$ at $T_{i-1}$.
\par 
In the case of OIS the underlying rate $R_x(T_{i-1},T_i) \equiv R_{on}(T_{i-1},T_i)$ is a compounded overnight rate with a daily tenor $(x=on)$ compounded across the time interval $\Brack{T_{i-1},T_i}$,
\begin{equation}
R_{on}(T_{i-1},T_i) 
:=\frac{1}{\tau(T_{i-1},T_i)}\left\{\prod_{k=1}^{n_i}\left[1+R_{on}(T_{i,k-1},T_{i,k}) \tau(T_{i,k-1}, T_{i,k})\right]-1\right\},
\label{eq: spot ON compounded rate}
\end{equation}
where for each coupon period $\Brack{T_{i-1},T_i}$ in the floating leg there is a nested sub-schedule including $n_i-1$ overnight fixing dates, 
\begin{gather}
\begin{split}
&\bm{T}_i=\Brace{T_{i,1},...,T_{i,k},...,T_{i,n_i}},\\
&T_{i,0}=T_{i-1},\quad T_{i,n_i}=T_{i},\\
&\bigcup_{k=1}^{n_i}\left(T_{i,k-1},T_{i,k}\right] = \left(T_{i-1},T_i\right],\\
&\sum_{k=1}^{n_i}\tau(T_{i,k-1},T_{i,k})] = \tau(T_{i-1},T_i).
\end{split}
\label{eq: OIS schedule}
\end{gather}
OIS forward rates $F^\alpha_{on,i}(t)$ are computed from OIS yield curve built from homogeneous market OIS quotes, where the overnight CSA rate $r_\alpha$ is equal to the underlying overnight rate $R_{on}$, a procedure called single-curve bootstrapping\footnote{since one single yield curve is involved, see \cite{AmeBia13} for a detailed discussion.}. 
Only in this case we may write the usual no-arbitrage expression of OIS forward rates
\begin{equation}
F_{\alpha,i}(t) := F^\alpha_{\alpha,i}(t) 
=\frac{1}{\tau_F(T_{i-1},T_i)}\left[\frac{P^\alpha(t;T_{i-1})}{P^\alpha(t;T_i)}-1\right],
\label{eq: OIS forward rate vs collateralised ZCBs}
\end{equation}
where, differently from eq. \eqref{eq: IRS forward}, now $P^\alpha(t;T_i)$ can be interpreted as the price of a ZCB issued by a counterparty under a CSA with collateral rate $r_\alpha$, as discussed in sec. \ref{Subsec: theory}.
\par
We stress that, in the pricing formulas above, we consider a general situation where the underlying floating rate $R_x(T_{i-1},T_i)$ can be different from the CSA rate $r_\alpha$. 
In particular, during the transition period, a \QuoteDouble{dirty discounting} situation is possible, where the same Swap could be cleared with a CCP using \euro STR-discounting, according to the CCP rules, or traded with a market counterparty using EONIA-discounting, according to the existing bilateral CSA. 
A similar situation could also happen whenever counterparties agree to transform bilateral CSA rates from EONIA to \euro STR$+8.5$ bps (to avoid NPV and collateral jumps and discussions on possible compensation schemes).

\subsection{XVA Pricing}
\label{Subsec: XVA}
In this section we focus on the Funding Valuation Adjustment (FVA) defined as the cost of financing a derivative position across its entire lifetime (also called Funding Cost Adjustment, FCA\footnote{We do not consider here the other funding-related adjustments, i.e. the Funding Benefit Adjustment (FBA) and the Margin Valuation Adjustment (MVA), but the discussion could be easily generalized to include also these XVA components.}),
\begin{align}
\begin{split}	
\FVA^{\alpha}_c(t) 
&= -\mathbb{E}^{\mathbb{Q}}\Brack{\int_{t}^{T}D^\alpha(t;u)\Brack{H^\alpha_c(u)}^+
	\mathds{1}_{\{\tau_c>u\}}\mathds{1}_{\{\tau_I>u\}}s^\alpha_I(u)\;du\Bigg|\mathcal{G}_t},\\
H^\alpha_c(u) &= \mathbb{E}^{\mathbb{Q}}_u\Brack{V^\alpha_{0,c}(u,T)}-C^\alpha_c(u),\\
s^\alpha_I(t) &= r_I(t)-r_\alpha(t),
\label{eq: fva}
\end{split}
\end{align}
where $H^\alpha_c(u)$ is the Institution's exposure at time $u>t$ relative to counterparty $c$, $V^\alpha_{0,c}(u,T)$ is the base value\footnote{we assume \QuoteDouble{risk free} closeout at the base value, without any adjustment} at time $u$ of any future cash flow exchanged with counterparty $c$ in the time interval $\Brack{u,T}$ (a.k.a. \QuoteDouble{mark to future}), $C^\alpha_c(u)$ is the $\alpha-$collateral amount (variation margin) posted by counterparty $c$, 
$s^\alpha_I(u)$ is the Institution's instantaneous funding spread, $\tau_x$ and $\mathds{1}_{\{\tau_x>u\}}$, $x\in\Brace{c,I}$ are the corresponding default times and survival indicators. 
The integral covers the entire lifetime of all trades with counterparty $c$ up to the last cash flow date $T$. 
The positive part in eq. \eqref{eq: fva} means that a positive exposure of the Institution $I$ with respect to counterparty $c$ carries a funding cost emerging from the corresponding hedging trades, either cleared with a CCP or subject to bilateral CSA, where the variation margin posted by the Institution\footnote{the Institution's exposure on hedging trades is specularly negative, so variation margin must be posted by the Institution to the CCP or to the hedging counterparty.} is funded and compensated at the instantaneous collateral rate $r_\alpha(t)$.  
\par 
Similar formulas for other XVAs\footnote{the most common XVAs are credit/debt valuation adjustments (CVA/DVA), which take into account the default risk of the two counterparties.} can be found in e.g. \cite{Green15,Greg20}. 
Survival probabilities used for XVAs are computed from default curves built from market CDS quotes using OIS yield curve for discounting. In the present context we assume that both market CDS and OIS used for default curve bootstrapping share the same CSA rates.
\par 
To the purposes of sec. \ref{Sec: XVA impacts}, we consider a simplified situation consisting in a single uncollateralized trade with one single deterministic future cash flow $C(T)$ received by the Institution $ I $ from counterparty $c$ at time $T$. 
In this case the Institution's exposure is always positive, hence the DVA for $I$ is zero.
We also assume that the counterparty $c$ is default-free, hence the CVA for $I$ is zero as well. 
As a consequence, only the FVA is non-null\footnote{Margin Valuation Adjustment (MVA) is also null since there is no initial margin, and we discard KVA} and reduces to
\begin{multline}
\text{FVA}^{\alpha}(t) =- \mathbb{E}^{\mathbb{Q}}\Brack{\int_{t}^{T}D^{\alpha}(t;u)\left[\mathbb{E}^{\mathbb{Q}}\left[D^\alpha(u,T)C(T)\Big|\mathcal{G}_u\right]\right]^+s_I(u)\mathds{1}_{\{\tau_I>u\}}\,du\Bigg|\mathcal{G}_t}\\
%&= -C(T)\mathbb{E}_{t}^{\mathbb{Q}}\Brack{\int_{t}^{T}D(t;u)P(u,T)s_I(u)\mathds{1}_{\{\tau_I>u\}}\,du}\\
%&= -C(T)\int_{t}^{T}\mathbb{E}_{t}^{\mathbb{Q}}\Brack{D(t;u)P (u,T)s_I(u)\mathds{1}_{\{\tau_I>u\}}}\,du\\
%&= -C(T)\int_{t}^{T}\mathbb{E}_{t}^{\mathbb{Q}}\Brack{D(t;u)P (u,T)} s_I(u)\mathbb{E}_{t}^\mathbb{Q}\Brack{\mathds{1}_{\{\tau_I>u\}}}\,du\\
= -C(T)\int_{t}^{T}\mathbb{E}^{\mathbb{Q}}
\Brack{e^{-\int_{t}^{u}r_\alpha(s)ds}\,\mathbb{E}^{\mathbb{Q}}
	\Brack{e^{-\int_{u}^{T}r_\alpha(s)ds}\Big|\mathcal{F}_u}\Big|\mathcal{F}_t} s_I(u) \mathbb{E}^\mathbb{Q}\Brack{\mathds{1}_{\{\tau_I>u\}}\Big|\mathcal{H}_t}\,du\\
= -C(T)\int_{t}^{T}\mathbb{E}^{\mathbb{Q}}\Brack{\mathbb{E}^{\mathbb{Q}}\Brack{D^{\alpha}(t;T)\big|\mathcal{F}_u}\big|\mathcal{F}_t} s_I(u) S_I(t,u)\,du\\
= -P^{\alpha}(t;T) C(T) \int_{t}^{T}s_I(u)S_I(t,u)\,du,
\end{multline}
where $S_{I}(t,T)=\mathbb{E}^{\mathbb{Q}}\Brack{\mathds{1}_{\{\tau_I>T\}}\big|\mathcal{H}_t}=\mathbb{Q}(\tau_I>T)$ is the survival probabiliy of $I$ until time $T$, evaluated in $t$ (see App. \ref{App: hazard rate}), and we assumed independence of credit and interest rate processes. 
Using the following relationship (see App. \ref{App: funding spread})
\begin{equation}
s_I(u)=-\dfrac{\partial_{u}S_{I}(t,u)}{S_{I}(t,u)}(1-\textit{Rec}_I),
\label{eq: inst spread  see app}
\end{equation}
and the following expression for the price of a risky Zero Coupon Bond issued by Institution $I$ with recovery rate $\textit{Rec}_I$ (see App. \ref{App: riskyZCB}),
\beq
P^{\alpha}_I(t;T)=P^{\alpha}(t;T)\Parenthesis{\textit{Rec}_I+(1-\textit{Rec}_I)S_{I}(t,T)},
\label{eq: risky ZCB}
\eeq
we obtain the simple expression for FVA,
\begin{equation}	
\begin{split}
\FVA^{\alpha}(t) 
&= - (1-\textit{Rec}_I)P^{\alpha}(t;T) C(T) \int_{t}^{T}\left[-\dfrac{\partial_{u}S_I(t,u)}{S_I(t,u)}\right]S_I(t,u)\,du\\
&= (1-\textit{Rec}_I) P^{\alpha}(t;T) C(T) \int_{t}^{T}\partial_{u}S_I(t,u)\,du\\
&= (1-\textit{Rec}_I) P^{\alpha}(t;T) C(T) [S_I(t,T)-1]\\
%&=C(T)P^{\alpha}(t;T)[(S_I(t,T)-1)(1-R)]\\
&= -[P^{\alpha}(t;T)-P^{\alpha}_I(t,T)]C(T)\\
&= -V^\alpha_0(t) + P^{\alpha}_I(t,T)C(T),
\end{split}
\label{eq: FVA simplified}
\end{equation}
and for the total fair value
\begin{equation}
V^\alpha(t) = V^\alpha_0(t) + \FVA^\alpha(t) = P^{\alpha}_I(t,T)C(T).
\label{eq: Fair Vale cum FVA}
\end{equation}
Eq. \eqref{eq: Fair Vale cum FVA} formalizes the intuition that including the funding cost amounts to discount at the funding rate. 
We stress that this simple result holds only under the simplified hypotheses described above.

%%%%%%%%%%%%%%%%%%%%%%%%%%%%%%%%%%%%%%%%%%%

\section{OIS Impacts}
\label{Sec: OIS impacts}
In this section we analyse the impacts of the transition from EONIA to \euro STR on EUR OIS instruments. 
Since an OIS depends on overnight rates both in the underlying index and in the discounting, i.e. collateral rate, we consider the impact of the transition on both sides. 
We analyse both the discrete (sec. \ref{Subsec: discrete}) and continuous compounding regimes (sec. \ref{Subsec: continuous}), relying on the basic theoretical concepts reported in sec. \ref{Subsec: OIS and IRS pricing}. 
The corresponding numerical results are discussed in sec. \ref{Subsec: OIS results}.

\subsection{Discrete Compounding}
\label{Subsec: discrete}
We start to price one single \euro STR OIS floating coupon (OISlet) using eq. \eqref{eq: IRS price} (shorting the notation of year fractions),
\begin{equation}
\OISlet_\textit{\euro ST}(t;T_{i-1},T_i,\alpha) 
= P^\alpha(t;T_i)\mathbb{E}^{Q^\alpha_{T_i}}\Brack{R_\textit{\euro ST}(T_{i-1},T_i)}\tau_{R,i},
\label{eq: oislet price}
\end{equation}
where, in principle, the collateral rate $r_\alpha$ is different from the underlying rate $R_\textit{\euro ST}$.
We now restrict our analysis to the particular case where the collateral rate $r_\alpha$ is either $R_\textit{EON}$ or $R_\textit{\euro ST}$, and we switch the underlying spot \euro STR compounded rate, given in eq. \eqref{eq: spot ON compounded rate}, to EONIA,
\begin{multline}
R_\textit{\euro ST}(T_{i-1},T_i) 
=\frac{1}{\tau_i}\left\{\prod_{k=1}^{n_i}\left[1+R_\textit{\euro ST}(T_{i,k-1},T_{i,k}) 
	\tau_{i,k}\right]-1\right\}\\
=\frac{1}{\tau_i}\left\{\prod_{k=1}^{n_i}\left[1+\Brack{R_\textit{EON}(T_{i,k-1},T_{i,k}) + \Delta} \tau_{i,k}\right]-1\right\}.
\label{eq: totalcomp}
\end{multline}
We show in App. \ref{App: OIS discrete} that the \euro STR OIS price and par rate are given by 
\begin{align}
\OIS_\textit{\euro ST}(t;\bm{T},\bm{S},K,\omega,\alpha)
&\approx \OIS_\textit{EON}(t;\bm{T},\bm{S},K,\omega,\alpha)
+ \omega N \sum_{i=\eta_R(t)}^n P^\alpha(t;T_i) \Sigma_{d,i}(\Delta) \tau_{R,i},
\label{eq: OIS price difference discrete}\\
R_\textit{\euro ST}(t;\textbf{T},\textbf{S},\alpha)
&\approx R_\textit{EON}(t;\textbf{T},\textbf{S},\alpha) + \delta_d(t;\Delta,\bm{T},\bm{S},\alpha),
\label{eq: OIS rate impact discrete}\\
\delta_d(t;\Delta,\bm{T},\bm{S},\alpha) 
&= \dfrac{\sum_{i=\eta_R(t)}^n P^\alpha(t;T_i)\Sigma_{d,i}(\Delta)\tau_{R,i}}{A(t;\bm{S},\alpha)},
\label{eq: OIS rate difference discrete}\\
\Sigma_{d,i}(\Delta)
&= F_\textit{\euro ST,i}(t) - F_\textit{EON,i}(t)\nonumber\\
&=\frac{\Delta}{\tau_{i}}
\Brack{\frac{P^\textit{EON}(t;T_{i-1})}{P^\textit{EON}(t;T_i)}
	\sum_{l=1}^{n_i} \tau_{i,l} \frac{P^\textit{EON}(t;T_{i,l})}{P^\textit{EON}(t;T_{i,l-1})}}.
%\new{riformulare i ratio in termini di tassi forward}\\
%&=\Delta\Brack{\Brack{F_{\textit{EON},i}(t)+1}
%	\sum_{l=1}^{n_i} \dfrac{1}{F_{\textit{EON},i,l}(t)+1}}
\label{eq: OIS Sigma discrete}
\end{align}
where $\delta_{d}(t;\Delta,\bm{T},\bm{S},\alpha)$ (\QuoteDouble{d} stands for \QuoteDouble{discrete}, to be compared with the continuous case in sec. \ref{Subsec: continuous}) is the spread between the EONIA and \euro STR OIS par rates due to the spread $\Delta$ on the overnight rates.
\par 
We notice that the forward rate spread $\Sigma_{d,i}(\Delta)$ in eq. \eqref{eq: OIS Sigma discrete} approaches the overnight rate spread $\Delta$ for vanishing rates, thanks to the additive property of the OIS year fractions in eq. \eqref{eq: OIS schedule}. 
As a consequence, also the par rate spread $\delta_d(t;\Delta,\bm{T},\bm{S},\alpha)$ in eq. \eqref{eq: OIS rate difference discrete} approaches $\Delta$, in particular when the fixed and floating schedules are the same (i.e. $\bm{T}=\bm{S}$).
\par 
We notice also that eqs. \eqref{eq: OIS price difference discrete}-\eqref{eq: OIS Sigma discrete} hold for any collateral rate which differs from EONIA by a constant spread. The generalization to a non-constant deterministic spread is straightforward.

%%%%%%%%

\subsection{Continuous Compounding}
\label{Subsec: continuous}
It is interesting to examine the limit case of continuous compounding where $\tau_{i,k}\rightarrow 0$ and the \euro STR spot overnight rate in eq. \eqref{eq: totalcomp} becomes
\begin{equation}\begin{split}
R_\textit{\euro ST}(T_{i-1},T_i)
&\rightarrow\dfrac{1}{\tau_{i}}
	\Brack{e^{\int_{T_{i-1}}^{T_i}r_{\textit{\euro ST}}(u)\,du}-1}\\
&=\dfrac{1}{\tau_{i}}
	\Brack{e^{\Delta\tau_{i}}e^{\int_{T_{i-1}}^{T_i}r_\textit{EON}(u)\,du}-1}.
\end{split}
\end{equation}
We show in App. \ref{App: OIS continuous} that in this case the \euro STR OIS price and par rate are given by 
\begin{align}
\OIS_\textit{\euro ST}(t;\bm{T},\bm{S},K,\omega,\alpha)
&\approx \OIS_\textit{EON}(t;\bm{T},\bm{S},K,\omega,\alpha)
+ \omega N \sum_{i=\eta_R(t)}^n P^\alpha(t;T_i) \Sigma_{c,i}(\Delta) \tau_{R,i},
\label{eq: OIS price difference continuous}\\
R_\textit{\euro ST}(t;\textbf{T},\textbf{S},\alpha)
&\approx R_\textit{EON}(t;\textbf{T},\textbf{S},\alpha) + \delta_c(t;\Delta,\bm{T},\bm{S},\alpha),
\label{eq: OIS rate impact continuous}\\
\delta_c(t;\Delta,\bm{T},\bm{S},\alpha) 
&= \dfrac{\sum_{i=\eta_R(t)}^n P^\alpha(t;T_i)\Sigma_{c,i}(\Delta)\tau_{R,i}}{A(t;\bm{S},\alpha)},
\label{eq: OIS rate spread continuous}\\
\Sigma_{c,i}(\Delta)
&= F_\textit{\euro ST,i}(t) - F_\textit{EON,i}(t)\nonumber\\
&=\dfrac{1}{\tau_{i}}\Brack{e^{\Delta\tau_{i}}-1}\dfrac{P^\textit{EON}(t;T_{i-1})}{P^\textit{EON}(t;T_{i})}.
\label{eq: OIS Sigma continuous}
\end{align}
%
%\begin{align}
%\OIS_\textit{\euro ST}(t;\bm{T},\bm{S},K,\omega,\alpha)
%&= \OIS_\textit{EON}(t;\bm{T},\bm{S},K,\omega,\alpha) \nonumber\\
%&+\omega N \sum_{i=\eta_R(t)}^n P^\alpha(t;T_i) \Sigma_{c,i}(\Delta)
%	\tau_{R,i},\\
%R_\textit{\euro ST}(t;\bm{T},\bm{S},\alpha) 
%&= R_\textit{EON}(t;\bm{T},\bm{S},\alpha) + \delta_c(t;\Delta,\bm{T},\bm{S},\alpha),
%\label{eq: OIS rate spread continuous}\\
%\delta_c(t;\Delta,\bm{T},\bm{S},\alpha)  
%&=\frac{1}{A(t;\bm{S},\alpha)} 		
%	\sum_{i=\eta_R(t)}^n P^{\alpha}(t;T_{i-1})\Sigma_{c,i}(\Delta),\\
%\Sigma_{c,i}(\Delta) &= \dfrac{1}{\tau_{R,i}}\Brack{e^{\Delta\tau_{R,i}}-1}\dfrac{P^\textit{EON}(t;T_{i-1})}{P^\textit{EON}(t;T_{i})}.
%\label{eq: spr}
%\end{align}
%We see that at first order in $ \Delta $ the discrete par OIS spread \eqref{eq: OIS rate difference discrete} tends to the continuous one,
%\begin{equation}
%\delta_d(t;\Delta,\bm{T},\bm{S},\alpha) \xrightarrow{\tau_{R,i,k}\rightarrow 0}
%\delta_c(t;\Delta,\bm{T},\bm{S},\alpha).
%\end{equation}
As for the discrete case, we notice that, for vanishing rates, $\Sigma_{c,i}(\Delta)\rightarrow\Delta$ and  $ \delta_c(t;\Delta,\bm{T},\bm{S},\alpha)  \rightarrow \Delta$.

\subsection{Numerical Results}
\label{Subsec: OIS results}
\begin{figure}[H]
	\centering
	\includegraphics[width=0.8\linewidth]{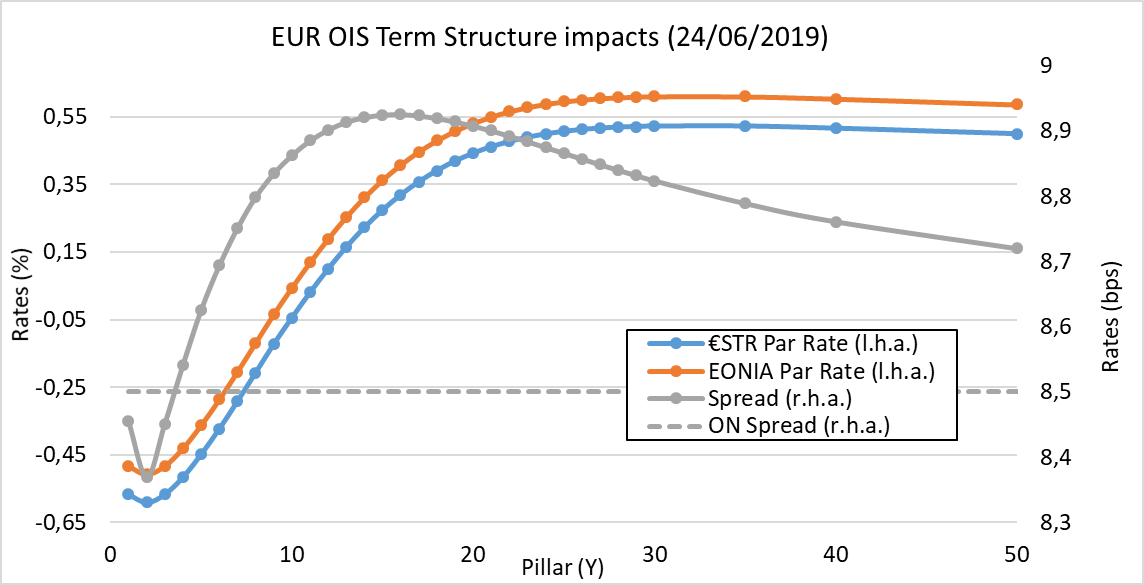}
	\caption{impact on OIS par rates. Left hand axis, continuous orange line: quoted EONIA OIS term structure as of 30/06/2020; continuous blue line: \euro STR OIS term structure built as described in sec. \ref{Subsec: discrete}. Right hand axis, continuous grey line: spread between EONIA and \euro STR par rates; dashed grey line: 8.5 basis point constant spread between EONIA and \euro STR overnight rates.}
	\label{Fig: OIS impacts 2019}
\end{figure}

\begin{figure}[H]
	\centering
	\includegraphics[width=0.8\linewidth]{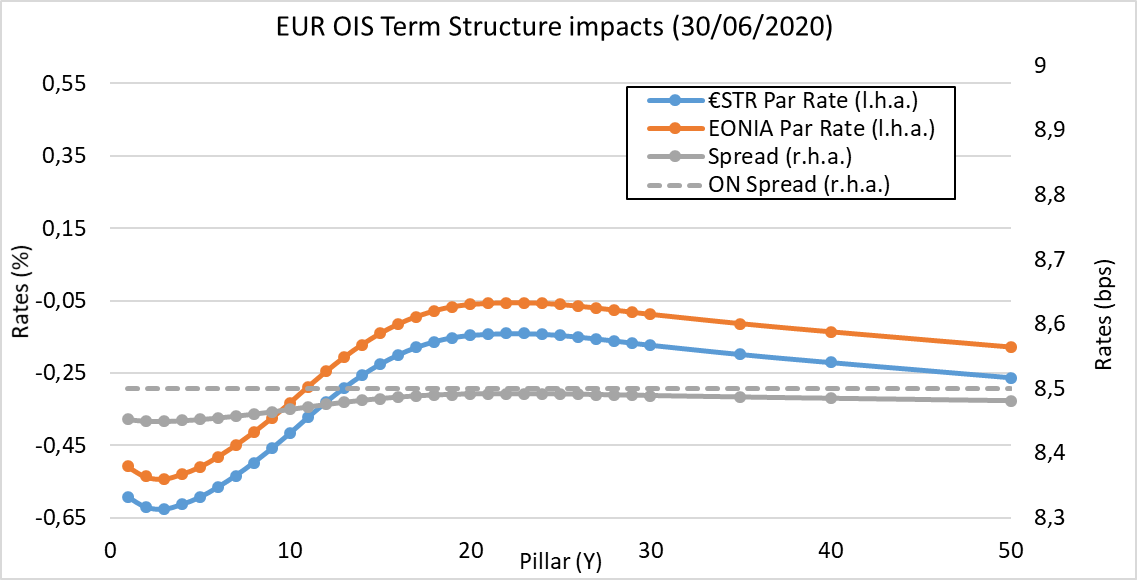}
	\caption{As in fig. \ref{Fig: OIS impacts 2019}, as of 30/06/2020.}
	\label{Fig: OIS impacts 2020}
\end{figure}

We analyse the impact of the transition from EONIA to \euro STR on OIS par rates comparing the quoted EONIA OIS term structure with the theoretical \euro STR OIS term structure built as described in sec. \ref{Subsec: discrete}, eq. \eqref{eq: OIS rate impact discrete}.
We show in figs. \ref{Fig: OIS impacts 2019} and \ref{Fig: OIS impacts 2020} the impact on two business dates (24/06/2019 and 30/06/2020), when the OTC derivative market quoted OISs assuming EONIA CSA, consistently with the approach of CCPs. Hence, we are using eq. \eqref{eq: OIS rate impact discrete} with overnight collateral rate $R_\alpha=$EONIA.
As expected, according to eqs. \eqref{eq: OIS rate difference discrete}-\eqref{eq: OIS rate impact discrete}, the spread between EONIA and \euro STR OIS par rates is not constant, but it depends on the level and shape of the term structure of EONIA OIS quotes. 
The difference with respect to the constant spread between EONIA and \euro STR overnight rates ($8.5$ bps) lies in the interval $[-0.2,+0.4]$ bps on the first date, an interval smaller than the typical market bid-ask spread of these instruments on the most liquid maturities (i.e. a couple of basis points).
On the second date the difference is reduced by one order of magnitude (RMSE = $ 0.03 $ bps versus RMSE = $ 0.34 $ bps), quite below the bid-ask spread, due to the significant lowering of the EONIA OIS term structure.
Part of this analysis is also reported in \cite{ECB19b}. This result is consistent with eqs. \eqref{eq: OIS rate difference discrete}-\eqref{eq: OIS rate impact discrete} and with the small value of $ 8.5 $ bps between \euro STR and EONIA overnight rates.
\par 
We conclude that the constant spread of $8.5$ bps between \euro STR and EONIA overnight rates does propagate to par OIS rates in a non-trivial way, but the residual distortion in the OIS term structure is quite small, depending on the level and shape of the market OIS par rates. 
As a consequence, one is allowed to bootstrap the \euro STR yield curve starting from EONIA OIS market quotes minus $8.5$ bps and, viceversa, bootstrap the EONIA yield curve starting from \euro STR OIS market quotes plus $8.5$ bps.

%%%%%%%%%%%%%%%%%%%%%%%%%%%%%%%%%%%%%%%%%%%

\section{IRS Impacts}
\label{Sec: IRS impacts}
In this section we analyse the impacts of the transition from EONIA to \euro STR on EUR IRS instruments, indexed to EURIBOR. 
Looking at the pricing formulas in sec. \ref{Subsec: OIS and IRS pricing} (eqs. \eqref{eq: IRS price}, \eqref{eq: IRS annuity} and \eqref{eq: IRS par rate}), we recognize that the switch has both a direct and an indirect effect, as also discussed in detail in \cite{ECB19b}.
The direct effect is observable on IRSs subject to EONIA-discounting (because of either clearing rules, CSA rules or internal policy), due to the simple switch to \euro STR-discounting, keeping constant the forward rates (\QuoteDouble{constant forward rates approach}).
The indirect effect is observable on EURIBOR forward rates because of the switch to \euro STR-discounting in the multi-curve bootstrapping procedure used to build the EURIBOR yield curves (see the discussion related to eq. \eqref{eq: IBOR forward rate vs risky ZCBs}), keeping constant the market IRS par rates (\QuoteDouble{constant par rates approach}).
We stress that the simultaneous switch of discounting curve both in pricing and in bootstrapping leads to a null impact on quoted par swap rates, since the bootstrapping procedure, by construction, fits the market quotes. 
\par 
In the following sections we analyse both impacts, using an \euro STR-discounting curve produced from the theoretical \euro STR quotes obtained by shifting the market EUR OIS quotes by $-8.5$ bps, as discussed in sec. \ref{Sec: OIS impacts}).

\subsection{Constant Par Rates Approach}
\label{Subsec: indirect impact}
The constant par rate approach assumes that market EUR IRS par rates are not affected by the transition. As a consequence the quantities in IRS par rate formula \eqref{eq: IRS par rate} behaves as follows,
\begin{equation}
\begin{split}
P^{\textit{EON}}(t;T_i)&\longrightarrow P^{\textit{\euro ST}}(t;T_i),\quad i=\eta_R(t),\dots,n,\\
P^{\textit{EON}}(t;S_j)&\longrightarrow P^{\textit{\euro ST}}(t;S_j),\quad j=\eta_K(t),\dots,m,\\
A(t;\textbf{S},\textit{EON})&\longrightarrow A(t;\textbf{S},\textit{\euro ST}),\\
F_{x,i}^\textit{EON}(t)&\longrightarrow F_{x,i}^\textit{\euro ST}(t),\quad i=\eta_R(t),\dots,n,\\
R_x(t;\textbf{T},\textbf{S},\textit{EON})&\longrightarrow R_x(t;\textbf{T},\textbf{S},\textit{\euro ST}) 
= R_x(t;\textbf{T},\textbf{S},\textit{EON}).\\
\end{split}
\end{equation}
\par
\begin{figure}[H]
	\centering
	\includegraphics[width=0.8\linewidth]{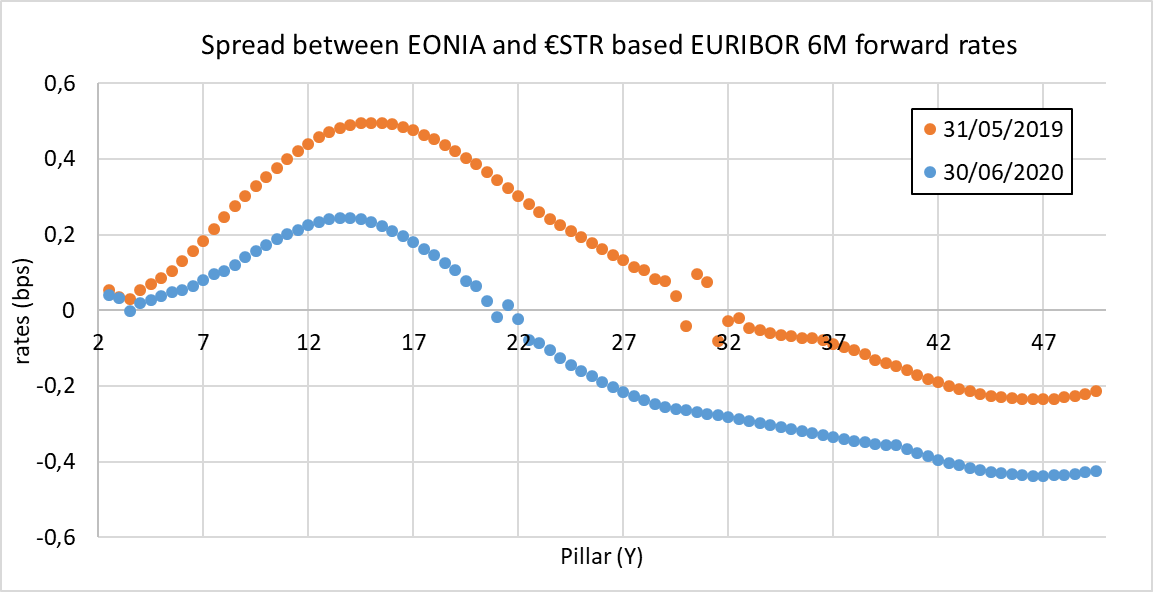}
	\caption{differences between EONIA and \euro STR based EURIBOR 6M forward rates $F_{6M,i}^\textit{EON}(t)-F_{6M,i}^\textit{\euro ST}(t)$ up to $T=50Y$, for $t=$31/5/2019 and $t=$30/6/2020. Minor irregularities are due to the interplay between yield curve shapes and interpolation effects.}
	\label{Fig: constant swp}
\end{figure}

 
We show in fig. \ref{Fig: constant swp} the differences between EONIA and \euro STR based EURIBOR 6M forward rates $F_{6M,i}^\textit{EON}(t)-F_{6M,i}^\textit{\euro ST}(t)$ up to $T=50Y$ for two different dates $t$ (total $100+100$ points). $F_{6M,i}^\textit{EON}(t)$ forward rates were computed from the EURIBOR 6M curve bootstrapped at time $t$ using EONIA-discounting, while $F_{6M,i}^\textit{\euro ST}(t)$ forward rates were computed from the EURIBOR 6M curve bootstrapped at time $t$ using \euro STR-discounting.
\noindent The differences result to be quite small, lying in the interval $[-0.4,+0.5]$ bps, and their term structure and global amount remain stable for the two dates examined (RMSE = $ 0.27 $ bps for both dates).
We conclude that the indirect impacts under the constant par rates approach are negligible.

\subsection{Constant Forward Rates Approach}
\label{Subsec: direct impact}
\begin{figure}[H]
	\centering
	\includegraphics[width=\linewidth]{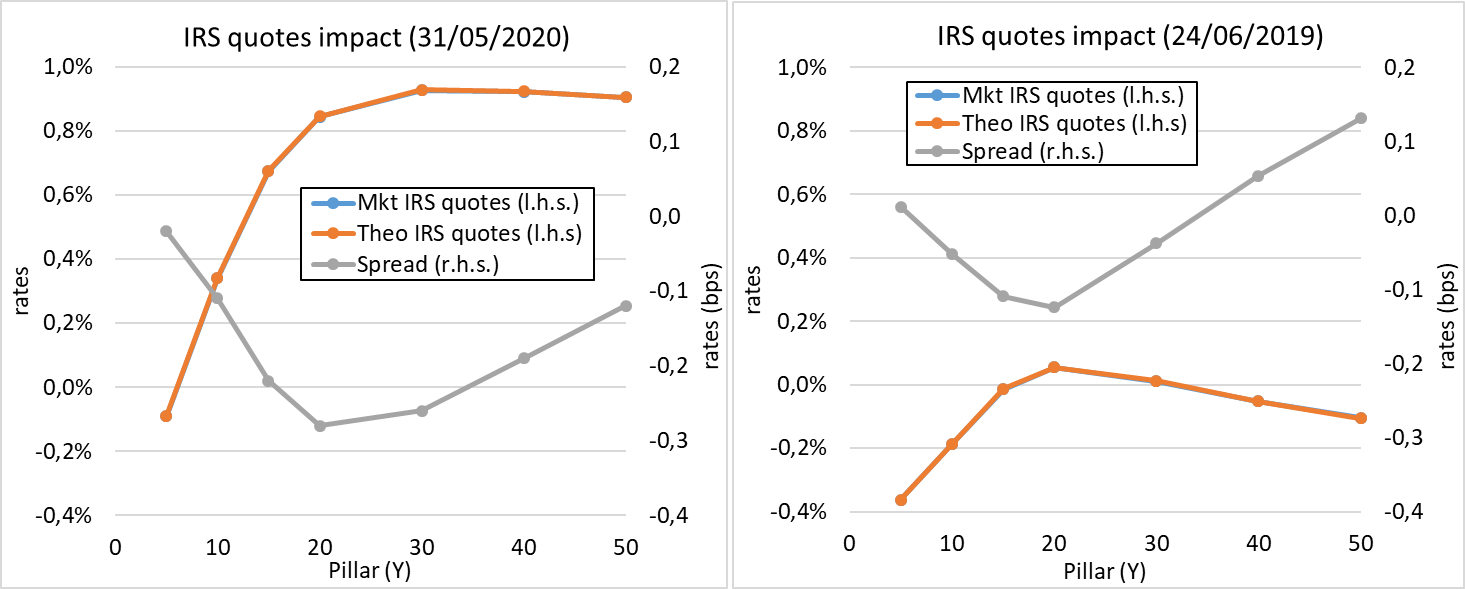}
	\caption{Discounting switch impacts on IRS par rates in the constant forward rates approach. Market IRS quotes (blue lines) refer to EURIBOR 6M IRS par rates (EONIA-discounting); theoretical quotes (orange lines) refer to EURIBOR 6M par rates in new \euro STR-discounting regime, obtained using EONIA based 6M EURIBOR forward rates; spreads (grey lines) refer to the differences between Market IRS quotes and theoretical quotes. }
	\label{Fig: IRS impacts}
\end{figure}

The constant forward rate approach assumes that EURIBOR forward rates are not affected by the transition. As a consequence the quantities in IRS par rate formula \eqref{eq: IRS par rate} behaves as follows,
\begin{equation}
\begin{split}
P^{\textit{EON}}(t;T_i) &\longrightarrow P^{\textit{\euro ST}}(t;T_i),\quad i=\eta_R(t),\dots,n,\\
P^{\textit{EON}}(t;S_j) &\longrightarrow P^{\textit{\euro ST}}(t;S_j),\quad j=\eta_K(t),\dots,m,\\
A(t;\textbf{S},\textit{EON}) &\longrightarrow A(t;\textbf{S},\textit{\euro ST}),\\
F_{x,i}^\textit{EON}(t) &\longrightarrow F_{x,i}^\textit{\euro ST}(t)=F_{x,i}^\textit{EON}(t),\quad i=\eta_R(t),\dots,n,\\
R_x(t;\textbf{T},\textbf{S},\textit{EON})&\longrightarrow R_x(t;\textbf{T},\textbf{S},\textit{\euro ST}).
\end{split}
\end{equation}
We show in Fig. \ref{Fig: IRS impacts} the impact of the transition on par swap rates. In particular, we show market quotes for EURIBOR 6M IRS par rates (EONIA discounted) versus theoretical EURIBOR 6M IRS par rates (\euro STR discounted) computed using eq. \eqref{eq: IRS par rate}, and their differences for two different dates. 
\noindent The differences result to be quite small, lying in the interval $ [-0.3,+0.1] $ bps; moreover, on 30/06/2020 the differences were sensibly reduced (RMSE = $0.09$ bps versus RMSE = $0.19$ bps). Part of this analysis is also reported in \cite{ECB19b}. 
\par 
Hence, also under the constant forward rate approach the impact of the discounting switch on IRS is negligible. 
\par 
Overall, we conclude that the transition from EONIA to \euro STR overnight rates affects both par IRS rates and implied forward rates in a negligible way, also considering multi-curve bootstrapping.
As a consequence, one may switch multi-curve bootstrapping of IBOR curves and IBOR-IRS pricing from EONIA to \euro STR quite safely.
Furthermore, the adoption of a \QuoteDouble{clean discounting} approach, where counterparties agree to switch bilateral CSAs from EONIA to \euro STR flat and to switch the discount curve accordingly, is theoretically sound and practically feasible.

\section{XVA Impacts}
\label{Sec: XVA impacts}
In this section we show that, assuming constant funding rates under the EONIA-\euro STR transition, the impact on the fair value $ V(t) $ of an uncollateralised trade is greatly reduced by a compensation effect between the impacts on the base value $ V_{0}(t)$ and on the $\FVA(t) $.

\subsection{Particular Case}
\label{Subsec: XVA impacts particular}
We will proof the proposition above in the simplified situation outlined in sec. \ref{Subsec: XVA}, using eq. \eqref{eq: FVA simplified}, which is referred to a single cash flow received at time $ T $ from a default free counterparty. According to eq. \eqref{eq: fair value} and \eqref{eq: FVA simplified}, the fair value of this trade is given, under EONIA-discounting, by
\beq
V^{\textit{EON}}(t)=V^{\textit{EON}}_{0}(t)+\FVA^{\textit{EON}}(t)=P^{\textit{EON}}_I(t;T)C(T).
\label{eq: fair value 2}
\eeq
Let's introduce the following zero rates,
\begin{equation}
\begin{split}
R(t;T,\textit{EON})&=-\dfrac{1}{\tau(t;T)}\log P^{\textit{EON}}(t;T),\\
R_I(t;T,\textit{EON})&=-\dfrac{1}{\tau(t;T)}\log P_{I}^{\textit{EON}}(t;T),\\
\mathcal{S}_{I}(t,T,\textit{EON}) &= R_I(t;T,\textit{EON}) - R(t;T,\textit{EON}),
\label{eq: zero rate}
\end{split}
\end{equation}
where $ \mathcal{S}_{I} $ is the funding zero spread. 
We assume here that the transition does not affect the Institution's absolute funding level, i.e. that bonds and other funded instruments issued by $I$ maintain constant prices across the transition. 
Hence, the Institution's funding (zero) rate $R_{I}(t;T,\textit{EON})$ remains constant across the transition, and the impact on the reference discount rate $R(t;T,\textit{EON})$ is absorbed by the corresponding impact on the funding spread $ \mathcal{S}_{I}(t,T,\textit{EON})$. 
As a consequence the quantities in eqs. \eqref{eq: zero rate} above behaves as follows,
\begin{align}
\begin{split}
R(t;T,\textit{EON}) &\longrightarrow R(t;T,\textit{\euro ST}),\\
R_{I}(t;T,\textit{EON}) &\longrightarrow R_I(t;T,\textit{\euro ST})=R_I(t;T,\textit{EON}),\\
P^{\textit{EON}}(t;T) &\longrightarrow P^{\textit{\euro ST}}(t;T),\\
P_I^{\textit{EON}}(t;T) &\longrightarrow P_I^{\textit{\euro ST}}(t;T)= P_I^{\textit{EON}}(t;T),\\
\mathcal{S}_I(t,T,\textit{EON}) &\longrightarrow  \mathcal{S}_{I}(t,T,\textit{\euro ST}).
\label{eq: XVA impacts}
\end{split}
\end{align} 
Notice that the transformation of $S_I$ in eq. \eqref{eq: XVA impacts} leads to
\begin{align}
\mathcal{S}_{I}(t,T,\textit{\euro ST}) 
&= R_I(t;T,\textit{\euro ST}) - R(t;T,\textit{\euro ST})\nonumber\\
&= \mathcal{S}_I(t,T,\textit{EON}) - \Brack{R(t;T,\textit{\euro ST})-R(t;T,\textit{EON})}
\approx \mathcal{S}_I(t,T,\textit{EON}) - \Delta,
\label{eq: funding spread impact}
\end{align} 
where the last equality is obtained from
\begin{multline}
R(t;T,\textit{\euro ST})-R(t;T,\textit{EON})=\dfrac{1}{\tau(t;T)}\log\dfrac{P^{\textit{EON}}(t;T)}{P^{\textit{\euro ST}}(t;T)}
=\dfrac{1}{\tau(t;T)}\log\dfrac{\mathbb{E}_t^{\mathbb{Q}}\Brack{e^{-\int_{t}^{T}r_{\textit{EON}}(u)\,du}}}
{\mathbb{E}_t^{\mathbb{Q}}\Brack{e^{-\int_{t}^{T}r_{\textit{\euro ST}}(u)\,du}}}\\
=\dfrac{1}{\tau(t;T)}\log\dfrac{\mathbb{E}_t^{\mathbb{Q}}\Brack{e^{-\int_{t}^{T}r_{\textit{EON}}(u)\,du}}}
{\mathbb{E}_t^{\mathbb{Q}}\Brack{e^{-\int_{t}^{T}\Brack{r_{\textit{EON}}(u)+\Delta}\,du}}}
=\dfrac{1}{\tau(t;T)}\log e^{\Delta(T-t) }
=\Delta \dfrac{(T-t)}{\tau(t;T)} 
\approx \Delta.
\label{eq: zero rate impact}
\end{multline}
The result in eq. \eqref{eq: zero rate impact} above is also reported in \cite{ECB19c}. We notice that possible different conventions for spread and zero rate year fractions could lead to an approximate equivalence. 
Using eq. \eqref{eq: XVA impacts}, the fair value in eq. \eqref{eq: fair value 2} becomes
\begin{equation}
\begin{split}	
V^{\textit{EON}}(t) = P_I^\textit{EON}(t;T)C(T) \longrightarrow 
P_I^\textit{\euro ST}(t;T)C(T) 
= P_I^\textit{EON}(t;T)C(T) 
= V^{\textit{EON}}(t),
\end{split}
\end{equation}
i.e., the fair value is constant across the transition.
\par 
The exact result proved above can be generalised to a collection of fixed positive cash flows (e.g. the fixed leg of receiver IRS), since the simplified FVA formula in eq. \eqref{eq: FVA simplified} still holds for each single cash flow.
\par 
We conclude that, under appropriate hypotheses, the impact of EONIA-\euro STR transition on the fair value $V$ of a trade is null, since the impact of the discounting switch on the base value $V_0$ is balanced by the corresponding impact on the FVA.

\subsection{General Case}
\label{Subsec: XVA impacts general}
In the general case of multiple stochastic cash flows with defaultable counterparties the XVAs formulas are more complex (see e.g. \cite{Green15,Greg20}) and the simple proof given in the previous sec. \ref{Subsec: XVA impacts particular} is no longer straightforward. 
In any case, guided by the simplified case, we may still leverage on the impacts analysed in sec. \ref{Sec: OIS impacts} to understand qualitatively the impacts of the EONIA to \euro STR transition, as follows\footnote{in our discussion we conventionally assume that CVA/DVA/FVA are negative/positive/negative adjustments, respectively.}.
\begin{description}
	\item[CVA:] it is a negative adjustment dependent on the Expected Positive Exposure (EPE) $\longrightarrow$ it is affected by a negative impact (CVA becomes more negative).
	\item[DVA:] it is a positive adjustment dependent on the Expected Negative Exposure (ENE) $\longrightarrow$ it is affected by a positive impact (DVA becomes more positive).
	\item[FVA:] it is a negative adjustment dependent on the Expected Positive Exposure (EPE) $\longrightarrow$ it is affected by a negative impact (FVA becomes more negative, as for CVA); furthermore, the FVA is also dependent on the funding spread $S_I$ $\longrightarrow$ it is affected by another negative impact.
\end{description}
We conclude that the impact of the transition on FVA is larger than for CVA and DVA, hence the cancellation effect between the FVA and the base value $V_0$ across the transition discussed in the previous sec. \ref{Subsec: XVA impacts particular} still exists. 
Notice that trades with negative exposure generate a positive impact. 
\par 
The discussion above actually depends on the XVAs management practice of Institutions, which is known to be quite diversified both in the selection of which XVAs are accounted at fair value and in the construction of own credit and funding curves (see e.g. \cite{XVARisk19}). We included CVA, DVA FVA (FCA actually) and we discarded the FBA, MVA and KVA, but this is only one possible choice. Hence, the exact conclusions regarding the impact of the EONIA to \euro STR transition should be carefully examined by each Institution according to its internal XVAs policy.

%%%%%%%%%%%%%%%%%%%%%%%%%%%%%%%%%%%%%%

\section{Conclusions}
\label{Sec: conclusion}
We have shown in detail how the transition from EONIA to \euro STR overnight rates affects the pricing of OIS, IRS and their associated XVAs, and what are the consequences on cleared, collateralised and non-collateralised linear financial instruments, both from a theoretical and numerical point of view.
\par 
In particular, we have shown that the constant spread of $8.5$ bps between \euro STR and EONIA overnight rates does propagate to par OIS rates in a non-trivial way, but the residual distortion in the OIS term structure is quite small, depending on the level and shape of the market OIS par rates. 
As a consequence, the differences between market quotes of \euro STR and EONIA OIS are expected to amount to $8.5$ bps, except in case of illiquidity problems, and one is allowed to safely bootstrap the \euro STR yield curve starting from EONIA OIS market quotes minus $8.5$ bps or, viceversa, bootstrap the EONIA yield curve starting from \euro STR OIS market quotes plus $8.5$ bps. These yield curves can be used to price EUR OIS indexed to EONIA or \euro STR. They can also be used as discounting curves for any kind of trade, consistently with either EONIA or \euro STR collateral.
\par 
Regarding IRS, we have shown that the transition from EONIA to \euro STR overnight rates affects both par IRS rates and implied EURIBOR forward rates in a negligible way, also considering multi-curve bootstrapping.
As a consequence, one may safely switch multi-curve bootstrapping of EURIBOR curves and EURIBOR IRS pricing from EONIA to \euro STR.
\par 
Finally, we have shown that the impact of the transition from EONIA to \euro STR on non-collateralised trades is greatly reduced, since changes in the base value tends to be counter-balanced by corresponding changes on Funding Value Adjustment (i.e. its negative component of Funding Cost Adjustment). 
\par 
The combination of the three findings above makes theoretically sound and safe the adoption of the \QuoteDouble{clean discounting} approach recommended by the ECB \cite{ECB19a}, EONIA-free and based on \euro STR only, to any trade: cleared, under bilateral CSA, non-collateralised, with a limited impact. 
Such conclusion is valid at least for linear IR derivatives, which cover most of the trading volume and existing positions in the EUR market. Analyses for non-linear IR derivatives can be found e.g. in \cite{Pit20,ECB20b}. 
Hence, after the EONIA to \euro STR switch performed on 27th July 2020 by Central Counterparties for cleared derivatives, the financial industry may safely proceed to switch both bilateral collateral agreements covering non-cleared EUR OTC derivatives, and, consistently, non-collateralised OTC derivatives including XVAs. 
Such EONIA-free pricing framework is essential for the complete elimination of EONIA before its discontinuation on 31st December 2021.

%--- Appendices
\begin{appendices}

\section{OIS Details}
\label{App: OIS}

\subsection{Discrete Compounding}
\label{App: OIS discrete}
We compute the \euro STR forward overnight compounded rate $F_\textit{\euro ST,i}(t)$ expected at time at time $t<T_i$ for the future OIS floating coupon period $\Brack{T_{i-1},T_i}$ as the expectation of the corresponding spot rate in eq. \eqref{eq: totalcomp} under the $T_i-$forward measure, using the forward measure equivalence $\mathbb{Q}^\textit{\euro ST}_{T_i} \sim \mathbb{Q}^\textit{EON}_{T_i}$ discussed in sec. \ref{Subsec: measure equiv},
\begin{equation}
\begin{split}
F_\textit{\euro ST,i}(t) 
&=\mathbb{E}_t^{\mathbb{Q}^\textit{\euro ST}_{T_i}}\Brack{R_\textit{\euro ST}(T_{i-1},T_i)}  
=\mathbb{E}_t^{\mathbb{Q}^\textit{EON}_{T_i}}\Brack{R_\textit{\euro ST}(T_{i-1},T_i)} \\
&=\frac{1}{\tau_{i}} \mathbb{E}_t^{\mathbb{Q}^\textit{EON}_{T_i}}
	\Brace{\prod_{k=1}^{n_i}
	\Brack{1+\Brack{R_\textit{EON}(T_{i,k-1},T_{i,k}) + \Delta} \tau_{i,k} } -1 }\\
&=\frac{1}{\tau_{i}} 
\Brace{\prod_{k=1}^{n_i}
	\Brack{1+ \Brack{\mathbb{E}_t^{\mathbb{Q}^\textit{EON}_{T_{i,k}}}
	\Brack{R_\textit{EON}(T_{i,k-1},T_{i,k})}+ \Delta} \tau_{i,k} } -1 }\\
&=\frac{1}{\tau_{i}} 
	\Brace{\prod_{k=1}^{n_i}
	\Brack{1+\Brack{F_\textit{EON,i,k}(t)+\Delta } \tau_{i,k}} -1 }\\
&=\frac{1}{\tau_{i}}
	\Brace{\prod_{k=1}^{n_i}
	\Brack{\frac{P^\textit{EON}(t;T_{i,k-1})}{P^\textit{EON}(t;T_{i,k})}+\Delta\,
	\tau_{i,k} }-1 },
\label{eq: RESTon}
\end{split}
\end{equation}
(see e.g. \cite{AmeBia13,Hen14} for detailed math), where $F_\textit{EON,i,k}(t)$ is the forward overnight rate observed at time $t$ for the future overnight period $\Brack{T_{i,k-1},T_{i,k}}$, a martingale under the $T_{i,k}-$ forward measure $\mathbb{Q}_{T_{i,k}}$ given in eq. \eqref{eq: IBOR forward rate vs risky ZCBs}.
%\begin{align}
%F_\textit{on,ij}^\textit{EON}(t) = F_\textit{on}^\textit{EON}(T_{i,k-1},T_{i,k})
%&=\mathbb{E}_{t}^{\mathbb{Q}_{T_{i,k}}} 	
%	\Brack{R_\textit{EON}(T_{i,k-1},T_{i,k})}\nonumber\\
%&=\frac{1}{\tau_F(T_{i,k-1},T_{i,k})}
%	\Brack{\frac{P^\textit{EON}(t;T_{i,k-1})}{P^\textit{EON}(t;T_{i,k})}-1 }.
%\label{eq: on fwd rate}
%\end{align}
In eq. \ref{eq: RESTon} we assumed $\tau_{F,i,k}=\tau_{i,k}\;\forall\; i,k$.
The product in the last row of eq. \eqref{eq: RESTon} may be manipulated as follows,
\begin{multline}
\prod_{k=1}^{n_i}
\Brace{\frac{P^\textit{EON}(t;T_{i,k-1})}{P^\textit{EON}(t;T_{i,k})}
	+\Delta\,\tau_{i,k} }\\
= \Brack{\frac{P^\textit{EON}(t;T_{i,0})}{P^\textit{EON}(t;T_{i,1})}
	+\Delta\,\tau_{i,1}}\cdot 
	\Brack{\frac{P^\textit{EON}(t;T_{i,1})}{P^\textit{EON}(t;T_{i,2})}
	+\Delta\,\tau_{i,2}}\cdots
	\Brack{\frac{P^\textit{EON}(t;T_{i,n_{i-1}})}{P^\textit{EON}(t;T_{i,n_i})}
	+\Delta\,\tau_{i,n_i}}\\
=\prod_{k=1}^{n_i} \frac{P^\textit{EON}(t;T_{i,k-1})}{P^\textit{EON}(t;T_{i,k})}
	+ \Delta \sum_{l=1}^{n_i} \tau_{i,l} \prod_{\substack{k=1\\k\neq l}}^{n_i}
	\frac{P^\textit{EON}(t;T_{i,k-1})}{P^\textit{EON}(t;T_{i,k})}
	+ f\Brack{\textit{O}(\Delta^m, m\geq2)},
\label{eq: product 2}
\end{multline}
where $f$ gathers the remaining terms of order $m\geq 2$ in the spread $\Delta$, which can be neglected. 
The second term in eq. \eqref{eq: product 2} can be written as follows,
\begin{align}
&\Delta \sum_{l=1}^{n_i} \tau_{i,l} \prod_{\substack{k=1\\k\neq l}}^{n_i}
\frac{P^\textit{EON}(t;T_{i,k-1})}{P^\textit{EON}(t;T_{i,k})}
= \Delta \sum_{l=1}^{n_i} \tau_{i,l}
	\frac{P^\textit{EON}(t;T_{i,l})}{P^\textit{EON}(t;T_{i,l-1})}
	\prod_{k=1}^{n_i}
	\frac{P^\textit{EON}(t;T_{i,k-1})}{P^\textit{EON}(t;T_{i,k})}.
\end{align}
Hence, considering the telescopic product
\begin{equation}
\prod_{k=1}^{n_i}
\frac{P^\textit{EON}(t;T_{i,k-1})}{P^\textit{EON}(t;T_{i,k})}
= \frac{P^\textit{EON}(t;T_{i,0})}{P^\textit{EON}(t;T_{i,n_i})}
= \frac{P^\textit{EON}(t;T_{i-1})}{P^\textit{EON}(t;T_i)},
\end{equation}
eq. \eqref{eq: RESTon} becomes
\begin{align}
F_\textit{\euro ST,i}(t) 
&\approx \frac{1}{\tau_{i}}
	\Brack{\frac{P^\textit{EON}(t;T_{i-1})}{P^\textit{EON}(t;T_i)}
	+ \Delta \frac{P^\textit{EON}(t;T_{i-1})}{P^\textit{EON}(t;T_i)}
	\sum_{l=1}^{n_i} \tau_{i,l} 
	\frac{P^\textit{EON}(t;T_{i,l})}{P^\textit{EON}(t;T_{i,l-1})}-1 }\nonumber \\
&= F_\textit{EON,i}(t)  + \Sigma_{d,i}(\Delta),
\label{eq: fwd ESTR vs EONIA}\\
\Sigma_{d,i}(\Delta)
&:=\frac{\Delta}{\tau_{i}}
	\Brack{\frac{P^\textit{EON}(t;T_{i-1})}{P^\textit{EON}(t;T_i)}
	\sum_{l=1}^{n_i} \tau_{i,l} \frac{P^\textit{EON}(t;T_{i,l})}{P^\textit{EON}(t;T_{i,l-1})}}.
\label{eq: sigma_i}
\end{align}
Eq. \eqref{eq: fwd ESTR vs EONIA} allows to express the \euro STR forward overnight compounded rate as the EONIA forward overnight compounded rate plus the spread in eq. \eqref{eq: sigma_i}. 
Hence, the price a \euro STR OIS is given, using eq. \eqref{eq: IRS price}, by
\begin{multline}
\OIS_\textit{\euro ST}(t;\bm{T},\bm{S},K,\omega,\alpha)
= \omega N \Brack{\sum_{i=\eta_R(t)}^{n} 
	P^\alpha(t;T_i) F_\textit{\euro ST,i}(t) \tau_{R,i} 
	- K A(t;\bm{S},\alpha)},\\
= \omega N \Brace{\sum_{i=\eta_R(t)}^{n} \Brack{P^\alpha(t;T_i) F_\textit{EON,i}(t) \tau_{R,i} 
	+ P^\alpha(t;T_i) \Sigma_{d,i}(\Delta) \tau_{R,i}} 
	- K A(t;\bm{S},\alpha)},\\
= \OIS_\textit{EON}(t;\bm{T},\bm{S},K,\omega,\alpha)
	+ \omega N \sum_{i=\eta_R(t)}^{n} P^\alpha(t;T_i) \Sigma_{d,i}(\Delta) \tau_{R,i}.
\end{multline}
The \euro ST OIS par rate is obtained by setting $\OIS_\textit{\euro ST}(t;\bm{T},\bm{S},K,\omega,\alpha)=0$ and solving for $K$. The result is given in eq. \eqref{eq: OIS rate difference discrete}.

\subsection{Continuous Compounding}
\label{App: OIS continuous}
If we consider the limit for $ \tau_{i,k} \rightarrow 0$, we can write \eqref{eq: totalcomp} as
\begin{equation}
\begin{split}
R_\textit{\euro ST}(T_{i-1},T_i)
&\approx\dfrac{1}{\tau_{i}}\left[e^{\int_{T_{i-1}}^{T_{i}}r_{\textit{\euro ST}}(u)d u}-1\right]\\
&=\dfrac{1}{\tau_{i}}\left[e^{\int_{T_{i-1}}^{T_{i}}[r_\textit{EON}(u) + \Delta]d u}-1\right]\\
&=\dfrac{1}{\tau_{i}}\left[e^{\Delta\tau_{i}}e^{\int_{T_{i-1}}^{T_{i}}r_\textit{EON}(u)d u}-1\right].
\end{split}
\end{equation}
As in the discrete case (eq. \eqref{eq: RESTon}), we compute the forward rate,
\begin{align}
F_\textit{\euro ST,i}(t) \nonumber
&:=\mathbb{E}_t^{\mathbb{Q}_{T_{i}}}\Brack{R_\textit{\euro ST}(T_{i-1},T_i)}\nonumber\\
&=\dfrac{1}{\tau_{i}}\Brace{e^{\Delta\tau_{i}}
	\mathbb{E}_t^{\mathbb{Q}_{T_{i}}}\left[e^{\int_{T_{i-1}}^{T_i}r_\textit{EON}(u)d u}\right]-1}\nonumber\\
&=\dfrac{1}{\tau_{i}}\Brace{e^{\Delta\tau_{i}}\dfrac{1}{P^\textit{EON}(t;T_i)}
	\mathbb{E}_t^{\mathbb{Q}}\left[e^{-\int_{t}^{T_{i}}r_\textit{EON}(u)d u}
	e^{\int_{T_{i-1}}^{T_i}r_\textit{EON}(u)d u}\right]-1}\nonumber\\
&=\dfrac{1}{\tau_{i}}\Brace{e^{\Delta\tau_{i}}\dfrac{1}{P^\textit{EON}(t;T_i)}\mathbb{E}_t^{\mathbb{Q}}\left[e^{-\int_{t}^{T_{i-1}}r_\textit{EON}(u)du}\right]-1}\nonumber\\
&=\dfrac{1}{\tau_{i}}\left[e^{\Delta\tau_{i}}\dfrac{P^\textit{EON}(t;T_{i-1})}{P^\textit{EON}(t;T_{i})}-1\right],\nonumber\\
%&=\dfrac{1}{\tau_{i}}\left[e^{\Delta\tau_{i}}\dfrac{P^\textit{EON}(t;T_{i-1})}{P^\textit{EON}(t;T_{i})}-1+\dfrac{P^\textit{EON}(t;T_{i-1})}{P^\textit{EON}(t;T_{i})}-\dfrac{P^\textit{EON}(t;T_{i-1})}{P^\textit{EON}(t;T_{i})}\right],\new{toglierei}\\
&=\dfrac{1}{\tau_{i}}\left[\dfrac{P^\textit{EON}(t;T_{i-1})}{P^\textit{EON}(t;T_{i})}-1\right]+
\dfrac{1}{\tau_{i}}\Parenthesis{e^{\Delta\tau_{i}}-1}\dfrac{P^\textit{EON}(t;T_{i-1})}{P^\textit{EON}(t;T_{i})}\nonumber\\
%&=F_\textit{EON,i}(t) +
%\dfrac{1}{\tau_{i}}\Brack{e^{\Delta\tau_{i}}-1}\dfrac{P^\textit{EON}(t;T_{i-1})}{P^\textit{EON}(t;T_{i})}\\
&=F_\textit{EON,i}(t) + \Sigma_{c,i} \label{eq: fwd ESTR vs EONIA c}\\
\Sigma_{c,i}&=\dfrac{1}{\tau_{i}}\dfrac{P^\textit{EON}(t;T_{i-1})}{P^\textit{EON}(t;T_{i})}\Parenthesis{e^{\Delta\tau_{i}}-1}
\label{eq: sigma_i c}
\end{align}
where we switched from $T_i-$forward measure $\mathbb{Q}_{T_i}$ to risk neutral measure $\mathbb{Q}$.
As in discrete case discussed in sec. \ref{App: OIS discrete}, eq. \eqref{eq: fwd ESTR vs EONIA c} allows to express the \euro STR forward overnight compounded rate as the EONIA forward overnight compounded rate plus the spread in eq. \eqref{eq: sigma_i c}. 
Hence, the price a \euro STR OIS is given, using eq. \eqref{eq: IRS price}, by
\begin{multline}
\OIS_\textit{\euro ST}(t;\bm{T},\bm{S},K,\omega,\alpha)
= \omega N \Brack{\sum_{i=\eta_R(t)}^{n} 
	P^\alpha(t;T_i) F_\textit{\euro ST,i}(t) \tau_{R,i} 
	- K A(t;\bm{S},\alpha)},\\
= \omega N \Brace{\sum_{i=\eta_R(t)}^{n} \Brack{P^\alpha(t;T_i) F_\textit{EON,i}(t) \tau_{R,i} 
		+ P^\alpha(t;T_i) \Sigma_{c,i}(\Delta) \tau_{R,i}} 
	- K A(t;\bm{S},\alpha)},\\
= \OIS_\textit{EON}(t;\bm{T},\bm{S},K,\omega,\alpha)
+ \omega N \sum_{i=\eta_R(t)}^{n} P^\alpha(t;T_i) \Sigma_{c,i}(\Delta) \tau_{R,i}.
\end{multline}
The \euro ST OIS par rate is obtained by setting $\OIS_\textit{\euro ST}(t;\bm{T},\bm{S},K,\omega,\alpha)=0$ and solving for $K$. The result is given in eq. \eqref{eq: OIS rate spread continuous}.

%%%%%%%%%%%%%%%%%

\section{Risky ZCB details}

\subsection{Hazard rate}
\label{App: hazard rate}
The hazard rate $ \gamma=\gamma(t) $ referred to the default time process $ \tau $ is defined as
\begin{equation}
\gamma(t)dt=\mathbb{Q}(t<\tau\leq t+dt|\tau>t)
\label{eq: hazard rate}
\end{equation}
that is the probability that  default occurs within the interval $ \mathopen{(}t,t+dt\mathclose{]} $, conditioned to the fact that the default has not already occurred at $ t $; using conditioned probability definition, we may write
\begin{equation}
\mathbb{Q}(t<\tau\leq t+dt)=\gamma(t)\mathbb{Q}(\tau>t)dt
\end{equation}
and then
\begin{equation}\begin{split}
\mathbb{Q}(t<\tau\leq t+dt|\tau>t)&=\mathbb{Q}(\textcolor{black}{\{\tau\leq t+dt\}}\cap\textcolor{black}{\{\tau>t\}})\\
&=\mathbb{Q}(\{\tau>t\}\setminus\{\tau>t+dt\})\\
&=\mathbb{Q}(\tau>t)-\mathbb{Q}(\tau>t+dt).
\end{split}\end{equation}
Setting $ \Gamma(t):=\mathbb{Q}(\tau>t) $ we get
\begin{gather}
\Gamma(t)-\Gamma(t+dt)=\gamma(t)\Gamma(t)dt\\
\Gamma'(t)=-\gamma(t)\Gamma(t).
\label{eq: gamma}
\end{gather}
Hence we obtain the following expression for the survival probability
\begin{gather}
\mathbb{Q}(\tau>t) = \Gamma(t) = e^{-\int_0^t\gamma(s)ds}.
\end{gather}
The derivation above assumes $\gamma$ deterministic. According to \cite{BriMer06} for $\gamma$ stochastic we have
\begin{gather}
\mathbb{Q}(\tau>t) 
:= \mathbb{E}^{\mathbb{Q}}\Brack{\mathds{1}_{\{\tau>T\}}\Big|\mathcal{H}_t}
= \mathbb{E}^\mathbb{Q}\Brack{e^{-\int_0^t\gamma(s)ds}\Big|\mathcal{H}_t}
:=S(0;t).
\label{eq: survival probability}
\end{gather}

\subsection{Risky Zero Coupon Bond}
\label{App: riskyZCB}
We define the risky zero coupon bond with no recovery rate as the bond issued by a defaultable issuer $I$ paying one unit of currency at maturity $T$ if the issuer $I$ has not defaulted before. The price at time $t<T$ is given by a generalization of eq. \eqref{eq: ZCBrf},
\begin{equation}
\begin{split}
P_I(t;T)=\mathbb{E}^\mathbb{Q}\Brack{D(t;T)\bm{1}_{\Brace{\tau_I>T}}|\mathcal{G}_t},
\end{split}
\end{equation}
where $\tau_{I}>t$ is the (stochastic) default time of the issuer. 
If interest rates and default time are independent of each other we obtain, using eq. \ref{eq: survival probability},
\begin{multline}
P_{I}(t;T) = 
\mathbb{E}^\mathbb{Q}\Brack{D(t;T)|\mathcal{F}_t}
\mathbb{E}^\mathbb{Q}\Brack{\bm{1}_{\Brace{\tau_I>T}}|\mathcal{H}_t}\\
= P(t;T)S_I(t;T)
=\mathbb{E}^{\mathbb{Q}}\left[e^{-\int_{t}^{T}\Brack{r(s)+\gamma_I(s)}ds}\Big|\mathcal{F}_t\right],
\label{eq: ZCBrr}
\end{multline}
where $S_{I}(t,T)$ is the survival probabiliy for $I$ until time $T$, evaluated in $t$, and $\gamma$ is the hazard rate defined in sec. \ref{App: hazard rate}. 
\par 
Let’s now consider the more realistic case where recovery rate is represented by a non-null process $\mathcal{R}_I$. 
It can be shown (see \cite{Bie13,Gra05}) that the price of a risky zero coupon bond becomes 
\begin{equation}\begin{split}
P_I(t;T)
&=\mathbb{E}_{t}^{\mathbb{Q}}\left[D(t;\tau_I)\mathcal{R}(\tau_{I})\mathds{1}_{\{\tau_{I}\leq T\}} 
+ D(t;T)\mathds{1}_{\{\tau_{I}>T\}} \right]\\
&=\mathbb{E}_t^{\mathbb{Q}}\left[\int_{t}^{T}\mathcal{R}_I(u)\gamma_I(u)e^{-\int_{t}^{u}\Brack{r(s)+\gamma_I(s)}ds}du\right] 
+ P_I(t;T).
\label{eq: risky zero coupon bond}
\end{split}\end{equation}
We adopt the recovery of treasury model (\cite{Bie13,Gra05}), such that 
\begin{equation}
\mathcal{R}_I(t) = \textit{Rec}_I\,P(t;T),
\label{eq: rt}
\end{equation}
where $ \textit{Rec}_I $ is a constant. Under such model the price of the risky zero coupon bond in eq. \eqref{eq: risky zero coupon bond} simplifies as
\begin{equation}\begin{split}
P_I(t;T)&= \textit{Rec}_I \mathbb{E}_{t}^{\mathbb{Q}}\left[\int_{t}^{T}P(u,T)\gamma_I(u)e^{-\int_{t}^{u}\Brack{r(s)+\gamma_I(s)}ds}du\right] + P_{I}(t;T)\\
&=\textit{Rec}_I \mathbb{E}_{t}^{\mathbb{Q}}\left[D(t;T)\int_{t}^{T}\gamma_I(u)e^{-\int_{t}^{u}\gamma_I(s)ds}du\right] + P_{I}(t;T)\\
&=\textit{Rec}_I \mathbb{E}_{t}^{\mathbb{Q}}\left[D(t;T)\left(1-e^{-\int_{t}^{T}\gamma_I(s)ds}\right)\right] + P_{I}(t;T)\\
&=\textit{Rec}_I \left[P(t;T)-P_{I}(t;T)\right]+P_{I}(t;T)\\
%&=\textit{Rec}_I P(t;T)+(1-\textit{Rec}_I)P_{I}(t;T)\\
%&=\textit{Rec}_I P(t;T)+(1-\textit{Rec}_I)P(t;T)S_{I}(t,T)\\
&=P(t;T)\Brack{\textit{Rec}_I+(1-\textit{Rec}_I)S_{I}(t,T)}.
\label{eq: ZCB rr}
\end{split}\end{equation}

\subsection{Funding Spread}
\label{App: funding spread}
The zero funding spread rate defined in eq. \eqref{eq: zero rate} can be written as
\begin{equation}
\mathcal{S}_I(t,T)=\dfrac{1}{T-t}\log\dfrac{P(t;T)}{P_I(t,T)}.
\label{eq: zero funding spread}
\end{equation}
Using eq. \eqref{eq: ZCB rr} above we obtain
%\begin{equation}
%\begin{split}
%P_I(t;T)
%%&=P(t;T)\left[\textit{Rec}_I+S_{I}(t,T)-\textit{Rec}_IS_{I}(t,T)+1-1\right]\\
%%&=P(t;T)\left[1-(1-\textit{Rec}_I)+S_{I}(t,T)(1-\textit{Rec}_I)\right]\\
%&=P(t;T)\Brace{1-(1-\textit{Rec}_I)\Brack{1-S_{I}(t,T)}}.
%\end{split}
%\end{equation}
%then
\begin{equation}
\begin{split}
\mathcal{S}_I(t,T)
&=\dfrac{1}{T-t}\log\dfrac{1}{1-(1-\textit{Rec}_I)\Brack{1-S_I(t,T)}}\\
&= -\dfrac{1}{T-t}\log\Brace{1-(1-\textit{Rec}_I)\Brack{1-S_I(t,T)}}\\
&= -\dfrac{1}{T-t}\log\Brack{1-1+S_I(t,T)+\textit{Rec}_I - \textit{Rec}_I S_I(t,T)}\\
&= -\dfrac{1}{T-t}\log\Brace{S_I(t,T)\Brack{1+\textit{Rec}_I\Parenthesis{-1+\dfrac{1}{S_{I}(t,T)}}}}.
\label{eq: spread zero}
\end{split}
\end{equation}
We now recall that the survival probability may be defined through the hazard rate (define in sec \ref{App: hazard rate})
\begin{equation}
S_I(t,T)=e^{-\int_t^T\gamma(s)ds}.
\label{eq: survival pbt}
\end{equation}
Inserting the survivial probability \eqref{eq: survival pbt} inside the spread rate formula \eqref{eq: spread zero}, we arrive to (see \cite{Jar97}, \cite{Bie13})
\begin{equation}\begin{split}
\mathcal{S}(t,T)&=\dfrac{\int_{t}^{T}\gamma(s)ds}{T-t} -\dfrac{1}{T-t}\log\Brack{1+\textit{Rec}_I\left(e^{\int_t^T\gamma(s)ds}-1\right)}.
\end{split}\end{equation}
In order to obtain the short spread rate, we apply the limit
\begin{equation}
\begin{split}
s_I(t) := \lim_{T\rightarrow t^{+}}\mathcal{S}(t,T)=\gamma(t)(1-\textit{Rec}_I).
\label{eq: inst spread}
\end{split}
\end{equation}
Using eq. \eqref{eq: gamma} we finally obtain
\begin{equation}
s_I(u)=-\dfrac{\partial_{u}S_I(t,u)}{S_I(t,u)}(1-\textit{Rec}_I).
\end{equation}

\end{appendices}

%--- Bibliography
\bibliographystyle{unsrt}
\bibliography{FinanceBibliography}

\end{document}